\documentclass[twocolumn,switch]{article}

\usepackage{preprint}
\usepackage{hyperref}
\usepackage[numbers,square]{natbib}

\usepackage{siunitx}
\usepackage[utf8]{inputenc}
\usepackage{longtable}
\usepackage{array}
\usepackage{multirow} 
\usepackage{multicol} 
\usepackage{booktabs} 
\usepackage{tabularx} 
\usepackage{amssymb} 
\usepackage{pifont} 
\usepackage{graphicx}
\usepackage{tabularray}
\usepackage{enumitem} 
\usepackage{float} 
\usepackage{xurl} 
\usepackage{subcaption}

\setlist[itemize]{nosep} 

\newcommand{\xmark}{\ding{55}} 

\hypersetup{
  colorlinks=false,       
  pdfborder={0 0 1},      
  linkbordercolor={1 0 0},  
  citebordercolor={0 1 0},  
  urlbordercolor={0 1 1}    
}

\usepackage[utf8]{inputenc}  
\usepackage[T1]{fontenc}
\usepackage{xcolor}
\usepackage{lineno}          
\usepackage{tikz}            
\usepackage{cleveref}        

\usepackage{newfloat}
\DeclareFloatingEnvironment[name={Supplementary Figure}]{suppfigure}
\usepackage{sidecap}
\sidecaptionvpos{figure}{c}

\usepackage{titlesec}
\titlespacing\section{0pt}{12pt plus 3pt minus 3pt}{1pt plus 1pt minus 1pt}
\titlespacing\subsection{0pt}{10pt plus 3pt minus 3pt}{1pt plus 1pt minus 1pt}
\titlespacing\subsubsection{0pt}{8pt plus 3pt minus 3pt}{1pt plus 1pt minus 1pt}

\definecolor{lime}{HTML}{A6CE39}
\DeclareRobustCommand{\orcidicon}{
  \begin{tikzpicture}
    \draw[lime, fill=lime] (0,0) circle [radius=0.16]
      node[white] {{\fontfamily{qag}\selectfont \tiny ID}};
    \draw[white, fill=white] (-0.0625,0.095) circle [radius=0.007];
  \end{tikzpicture}
  \hspace{-2mm}
}

\title{Privacy vs. Profit: The Impact of Google's Manifest Version 3 (MV3) Update on Ad Blocker Effectiveness}

\usepackage{authblk}

\author[1]{Karlo Lukic\thanks{\texttt{lukic@wiwi.uni-frankfurt.de}} \href{https://orcid.org/0000-0003-2745-6473}{\orcidicon}}
\author[1]{Lazaros Papadopoulos\thanks{\texttt{papadopoulos@wiwi.uni-frankfurt.de}}\href{https://orcid.org/0009-0008-8011-2913}{\orcidicon}}

\affil[1]{Goethe University Frankfurt}

\begin{document}

\twocolumn[\begin{@twocolumnfalse}

\maketitle


\noindent\begin{center}
\begin{minipage}{0.96\linewidth}\centering
\small\textbf{Note.} This is the \textit{camera-ready} version accepted to
\textit{Proceedings on Privacy Enhancing Technologies (PoPETs), Volume 2026, Issue 1}.
The official version with page numbers and DOI will appear on the PoPETs website when the 2026(1) issue is released.

\vspace{0.4em}

\small\textbf{Please cite as:} \\
K. Lukic and L. Papadopoulos. ``Privacy vs. Profit: Impact of Google’s MV3 on
Ad-Blocker Effectiveness.'' \textit{Proceedings on Privacy Enhancing Technologies}, 2026(1) (forthcoming).
\end{minipage}
\end{center}
\vspace{0.5em}

\begin{abstract}

Google’s recent update to the manifest file for Chrome browser extensions, transitioning from manifest version 2 (MV2) to manifest version 3 (MV3), has raised concerns among users and ad blocker providers, who worry that the new restrictions, notably the shift from the powerful \textit{WebRequest} API to the more restrictive \textit{DeclarativeNetRequest} API, might reduce ad blocker effectiveness. Because ad blockers play a vital role for millions of users seeking a more private and ad-free browsing experience, this study empirically investigates how the MV3 update affects their ability to block ads and trackers.

Through a browser-based experiment conducted across multiple samples of ad-supported websites, we compare the MV3 to MV2 instances of four widely used ad blockers. Our results reveal no statistically significant reduction in ad-blocking or anti-tracking effectiveness for MV3 ad blockers compared to their MV2 counterparts, and in some cases, MV3 instances even exhibit slight improvements in blocking trackers. These findings are reassuring for users, indicating that the MV3 instances of popular ad blockers continue to provide effective protection against intrusive ads and privacy-infringing trackers. While some uncertainties remain, ad blocker providers appear to have successfully navigated the MV3 update, finding solutions that maintain the core functionality of their extensions.

\end{abstract}

\keywords{manifest version 3 (MV3), ad blocker, online advertising, online tracking}

\vspace{0.5cm}

\end{@twocolumnfalse}]


\section{Introduction}

Browser extensions are small software programs that can improve the capabilities of a web browser. Among them, ad blockers stand out. These web privacy tools help users block advertisements (ads) and website trackers, improving their browsing experience and level of privacy \cite{Frik2020, Gervais2017, Singh2009}. Around 31.5\% of internet users worldwide use an ad blocker, indicating widespread use and popularity \cite{Dean2024}.

A crucial component of browser extensions (hereafter, "extensions") is the manifest file, which outlines the extension’s capabilities \cite{Frisbie2023, Picazo2022}. In simple terms, this file acts as an instruction manual for the extension. It specifies what the extension can do and how it can interact with the user’s browser and visited websites, outlining its capabilities.

In December 2020, Google introduced an update from manifest version 2 (MV2) to manifest version 3 (MV3; \cite{Li2020}). Google positions this update as a strategy to improve the privacy, security, and performance of Chrome browser users who rely on extensions \cite{Cronin2019, Wagner2018}. However, other stakeholders, particularly users and ad blocker providers, view this update as a strategy to reduce the effectiveness of ad blockers, potentially increasing the ad revenue of Google and website publishers \cite{Barnett2021}.

The MV3 update restricts the ability of ad blockers to intercept and modify network requests, which has raised concerns about their possibly reduced effectiveness in blocking ads and trackers. Nevertheless, since the MV3 update has been announced, ad blocker providers have worked intensely to close the gap and provide a seemingly identical experience under the new restrictions. Despite these efforts users remain concerned about the impact of the MV3 update on ad-blocking effectiveness and consequently their privacy \cite{ArsTechnica2023, Verge2024}.

These user concerns show that Google’s underlying motivation for the update remains opaque, primarily because its internal decision-making processes and strategic incentives are not fully visible to the public. The MV3 update might deliberately weaken the ad-blocking and anti-tracking effectiveness of ad blockers, thereby consolidating Google’s control over user data and increasing its ad revenue. This outcome would align with patterns in digital advertising, where companies actively work to bypass user interventions, like ad blockers, to protect and expand their monetization potential (e.g., \cite{Acquisti2015}).

Alternatively, the MV3 update could represent a strategic move toward increased user privacy. Previous research has shown that only a small number of blocking rules are effectively used by ad blockers \cite{Snyder2020}, leaving much room for optimization. Further, by reducing potential avenues for intrusive data collection, Google might aim to strengthen user trust, a factor that research in privacy economics (e.g., \cite{Acquisti2015}) suggests can lead to more user engagement. Such research has shown that enhanced privacy protections are often correlated with increased platform loyalty, as users tend to favor services that safeguard their data.

As it is both technically and economically unclear whether the MV3 update reduces ad blocker effectiveness, we empirically investigate its effects on ad blockers. Therefore, this study measures the impact of Google’s MV3 update on ad blocker effectiveness, focusing on their ad-blocking and anti-tracking effectiveness.

Specifically, the study addresses the following research questions:

\begin{itemize}

\item \textbf{RQ1}: Does the effectiveness of MV3 ad blockers in blocking ads and trackers change when compared to that of their MV2 counterparts?

\item \textbf{RQ2}: Does the effectiveness of individual MV3 ad blockers differ from their own MV2 counterparts by the same ad blocker provider?

\item \textbf{RQ3}: Does the effectiveness of using multiple MV3 ad blockers simultaneously differ from using individual MV3 ad blockers?

\end{itemize}

The first research question aims to assess whether the transition from MV2 to MV3 impacts the effectiveness of MV3 ad blockers overall. It focuses on whether MV3 ad blockers are more restricted in blocking ads and trackers compared to their MV2 counterparts.

The second research question builds on this comparison by examining the effectiveness of individual MV3 ad blockers compared to their MV2 counterparts. Since most ad blocker providers state that their MV3 instances use either a reduced or modified set of blocking rules derived from their MV2 instances \cite{Adguard2024, uBlockOrigin2024, Adblock2024, Stands2024}, MV2 instances of ad blockers serve as natural benchmarks. By directly comparing each MV3 ad blocker to its MV2 counterpart, including Adblock Plus, AdGuard, Stands, and uBlock Origin, we assess how ad blocker providers have adapted to the MV3 update. This comparison is particularly relevant for users transitioning to MV3, as it provides insights into whether some MV3 ad blockers remain as effective as their MV2 counterparts, while others may exhibit reduced effectiveness. Thus, the second research question aims to help users choose the most effective MV3 ad blocker amid the MV2 to MV3 update. It also evaluates whether different ad blocker providers have adjusted differently to the MV3 update \cite{Adblock2024, Brinkmann2022, Ghostery2023, Meshkov2023, Orlova2023}.

The third research question addresses whether using multiple MV3 ad blockers simultaneously impacts ad-blocking and anti-tracking effectiveness compared to using individual MV3 ad blockers. This research question is relevant because it addresses concerns expressed by ad blocker providers about the potentially reduced effectiveness of the MV3 ad blocker when it is used simultaneously with multiple other MV3 ad blockers \cite{Seregin2022}. This concern is substantiated by the global rule limit introduced with the MV3 update. The third research question also measures the effectiveness of MV3 ad blockers for a privacy-aware user who uses multiple MV3 ad blockers simultaneously.

Thus, this study explores the impact of Google’s update from MV2 to MV3 on the effectiveness of ad blockers in blocking ads and trackers. Through a browser-based experiment, we compare MV3 ad blockers to their MV2 counterparts and examine the effect of using multiple MV3 ad blockers simultaneously. We initiate the browser-based experiment on a main sample of 924 websites across five separate measurement runs, and two additional samples of 191 websites and 185 websites stratified by website employee count and popularity rank, respectively. In addition, we run the browser-based experiment using Chrome and Firefox on our main sample of 924 websites, to ensure the cross-browser robustness of our results. We complement these experiments with additional analyses including alternative effectiveness metrics, assessments of early MV3 implementations, and detailed visual inspections of website screenshots that reinforce the robustness and generalizability of our findings.

Previous studies have laid the groundwork by evaluating the effectiveness of MV2 ad blockers, highlighting variations in their ad-blocking and anti-tracking effectiveness \cite{Borgolte2020, Garimella2017, Merzdovnik2017, Wills2016}. Building on existing literature, this study takes a step forward by empirically comparing MV3 to MV2 ad blockers. This exploration fills a critical gap in the existing literature. It contributes to understanding the evolution of ad-blocking tools and their implications for online privacy and security in the wake of Google’s MV3 update.

In summary, the main contributions of this study are:

\begin{itemize}

\item \textbf{Empirical comparison between MV3 and MV2 groups of ad blockers}: This study presents the first empirical analysis to evaluate the effectiveness of MV3 ad blockers compared to their MV2 counterparts. The results reveal no statistically significant reduction in ad-blocking effectiveness and a statistically significant increase in anti‑tracking effectiveness, with the MV3 group blocking about 1.8 more trackers per website on average than the MV2 group of ad blockers.

\item \textbf{Variability among individual MV3 ad blockers}: When comparing individual ad blockers, we find no significant differences between MV3 and MV2 instances for AdGuard and uBlock; however, Adblock Plus MV3 blocks about 1.9 more trackers than its MV2 counterpart (21.5\%), and Stands MV3 blocks about 5.2 more trackers (45.9\%) than its MV2 counterpart.

\item \textbf{Enhanced anti-tracking effectiveness with multiple MV3 ad blockers}: Using multiple MV3 ad blockers simultaneously does not impact their ad-blocking effectiveness. However, it significantly enhances anti-tracking effectiveness compared to using some MV3 ad blockers alone. Specifically, the combination blocks about 10.3 more trackers (95.3\%) than Adblock Plus MV3 alone, 4.4 more (26.6\%) than AdGuard MV3 alone, and 4.6 more (27.8\%) than Stands MV3 alone, with no difference relative to uBlock MV3. This improvement in anti-tracking effectiveness is driven by the inclusion of uBlock MV3 in the combination, suggesting a strategic advantage in combining certain MV3 ad blockers to improve users’ online privacy. This result alleviates concerns of ad blocker providers that using multiple MV3 ad blockers simultaneously reduces their effectiveness.

\end{itemize}

In additional tests, our findings remain robust under a variety of conditions. Results are consistent across different website samples, alternative effectiveness metrics, and over time. Moreover, cross-browser experiments yield comparable outcomes, and visual inspections of screenshots confirm that ad blockers operate effectively without significant ad flickering or loss of functionality. However, we do observe that MV3 ad blockers tend to produce a slightly less visually appealing browsing experience than their MV2 counterparts, primarily due to increased visibility of cosmetic placeholders.

The following sections of the research paper outline the background for the manifest file and MV2 ad blocker (Section~\ref{sec:background}), an overview of Google’s MV3 update (Section~\ref{sec:overview-of-google-mv3-update}), review related work (Section~\ref{sec:related-work}), present the study’s methodology and results (Sections~\ref{sec:measurement-setup} and \ref{sec:results}), and close with the limitations and conclusion (Sections~\ref{sec:limitations-and-ethics}, and \ref{sec:conclusion}).

\section{Background}\label{sec:background}

\subsection{Manifest File}\label{sec:manifest-file}

The manifest (\texttt{manifest.json}) file is a crucial part of any extension, acting as a communication bridge between the extension’s developer, the extension, and the user’s browser. This file serves as an instruction manual, translating the extension developer’s coding decisions into specific instructions that dictate how the browser should interpret the extension and act upon it for the user \cite{Frisbie2023, Picazo2022}.

At its core, the manifest file defines the extension’s capabilities and outlines how it will interact with the browser and across user-visited websites. It does that by defining "permissions" or "Application Programming Interfaces (APIs)"—terms that reflect the perspectives of the user and the developer, respectively \cite{Picazo2022}. APIs are specific authorizations the extension requires to function within the browser, such as accessing and changing content on all user-visited websites. Moreover, the manifest file can define other components of an extension, like \textit{background pages}, essentially scripts running "quietly" in the browser’s background \cite{Frisbie2023}. These scripts enable the extension to continuously perform tasks for the user without the need for the user to interact with the extension.

\begin{figure}[ht]
  \centering
  \setlength{\fboxsep}{1pt}
  \fbox{%
    \includegraphics[width=\dimexpr\linewidth-2\fboxrule-2\fboxsep\relax]{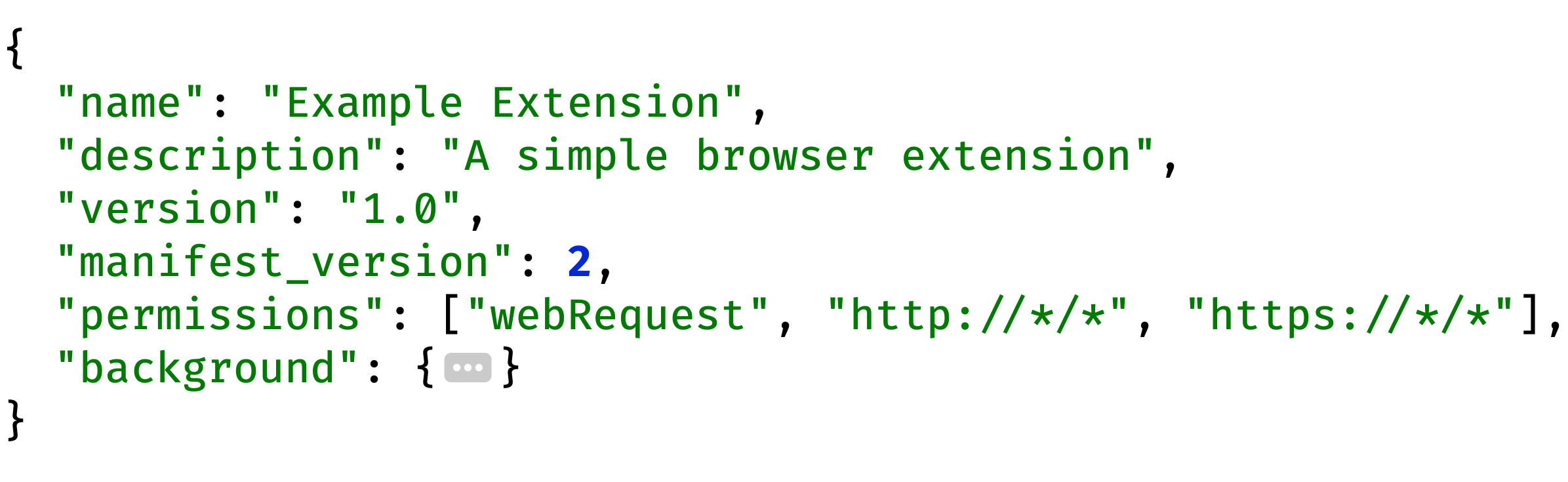}%
  }
  \caption{Illustration of the manifest (\texttt{manifest.json}) file for an extension.}
  \label{fig:figure-1}
\end{figure}

Figure \ref{fig:figure-1} illustrates a simplified example of a manifest file for an extension.

In Figure \ref{fig:figure-1}, the manifest file defines not only the basic details of the extension, such as its name, description, version, and manifest version, but also its need for the \textit{WebRequest} API and the inclusion of \textit{background pages} through the \textit{background} component.

The \textit{WebRequest} permission/API is vital for the extension to monitor, analyze, and modify the browser’s Hypertext Transfer Protocol (HTTP) requests: essentially, requests a user’s browser makes when visiting websites \cite{Frisbie2023}. While not specifying a particular file, the background component signals that the extension developer has designed the extension to run in the background, ensuring its continuous, uninterrupted core functionality.

When users install this extension, their browser reads and interprets its manifest file to understand its required permissions and other components, like the \textit{background} component. Thus, installing the "Example Extension," for instance, the user grants it the \textit{WebRequest} permission and allows it to run in the background of their browser.

This installation process represents the role of the manifest file as a communication bridge between the extension’s developer, the extension, and the user’s browser. It captures and conveys the developer’s intent in a format the browser can understand and act upon for the user, ensuring that the extension operates as designed within the scope of granted APIs.

Thus, the manifest file is a crucial part of an extension. By defining APIs and other components, the manifest file empowers extensions to perform specific, intended actions in the browser while ensuring these actions align with the extension developer’s intentions and the user’s expectations for privacy and security.

\subsection{Manifest Version 2 (MV2) Ad Blocker}\label{sec:mv2-ad-blocker}

An ad blocker is an extension that blocks ads and trackers from appearing on the websites users visit. Thus, an ad blocker has two core functions for the user: ad-blocking and anti-tracking. The ad blocker fulfills those two functions by monitoring and blocking HTTP requests.

In simple terms, the ad blocker contains a list of blocking rules. As the user’s browser makes each HTTP request, the ad blocker checks it against that list of blocking rules and blocks the request if it finds a match. Under Google’s MV2, ad blockers use the \textit{WebRequest} API and \textit{background pages} to monitor real-time HTTP requests between a user’s browser and the website they visit.

For instance, when users visit a website like \texttt{cnn.com}, their browser sends HTTP requests to servers owned by \texttt{cnn.com} (first party) and its partners (third parties; \cite{Mayer2012}). Using the \textit{WebRequest} API, the MV2 ad blocker can check these requests against a vast list of blocking rules, allowing it to block those HTTP requests directed toward third parties, typically including requests to display ads and load trackers. By doing so, MV2 ad blockers can block ads and trackers in real-time, ensuring users an ad-free browsing experience and increased privacy.

Users can install extensions and ad blockers from a browser provider’s extension store, like Google’s Chrome Web Store (CWS). Likewise, developers of extensions, specifically ad blocker providers, can publish their extensions in the browser provider’s extension store.

\section{Overview of Google's Manifest Version 3 (MV3) Update}\label{sec:overview-of-google-mv3-update}

\subsection{Description of Google's MV3 Update From Two Opposing Perspectives}\label{sec:description-of-mv3-from-2-perspectives}

Google announced the MV3 update in November 2018 and made it available to extension developers in beta form in December 2020 \cite{Li2020, Wagner2018}. In January 2022, Google stopped allowing developers to add new MV2 extensions to the CWS, signaling the start of the MV2 phase out \cite{Li2021}. In late 2024, Google warned developers that it intended to complete the MV2 phase out by July 2025 \cite{Google2024}.

With the MV3 update, Google claims to enhance user privacy, security, and potentially, Chrome browser performance \cite{Li2020}. Its stated aim is driven primarily by concerns about extensions, particularly because a significant percentage of them have turned malicious in the past \cite{Cronin2019, Wagner2018}. Some examples of extensions turning malicious include extensions stealing data from users, extensions injecting ads into users’ webpages or redirecting them to webpages with ads (a practice known as "malvertising"), and extensions making users’ browsers mine cryptocurrencies (a practice known as "cryptojacking"; \cite{Kaya2020, Newman2018}).

By introducing the MV3 update, Google states that it seeks to reduce the power of extensions and enhance Chrome browser performance, thereby restricting the potential misuse of user data by malicious extensions. The two critical technical changes in the MV3 update reflect Google’s stated effort to enhance the security and privacy of extensions \cite{Vincent2019}. The first change replaces the \textit{WebRequest} API, which allows ad blockers to monitor and block ads and trackers in real-time, with the \textit{DeclarativeNetRequest} API. This change restricts ad blockers to operating based on a predefined set of up to 30,000 blocking rules. That is less than half the number of rules in EasyList\footnote{\href{https://easylist.to}{https://easylist.to}}, the base filter set used by many extensions (e.g., uBlock Origin), and significantly restricts their processing pipeline. However, empirical analyses indicate that this constraint might not necessarily diminish ad blocker effectiveness \cite{Snyder2020, Bottger2025}.

The second change, the transition from \textit{background pages} to \textit{service workers}, intends to improve extension performance by managing background tasks more efficiently. A \textit{service worker}, unlike a more persistent \textit{background page}, spins up only when an extension needs it (i.e., it is event-driven) and is torn down after 30 seconds of idle time. Because the \textit{service worker} runs off the main thread and does not linger in memory, Chrome can reclaim RAM and CPU that dormant extensions once consumed, giving users a leaner, faster browser. The shorter life cycle might also reduce the window in which malicious or buggy code could abuse persistent privileges, so the change aims to strengthen security by design \cite{ChromeDevServiceWorker2023}.

Moreover, the MV3 update represents an effort towards unifying extensions across browsers. By collaborating with other browser providers and participating in initiatives like the W3C Browser Extensions Community Group \cite{W3C2023}, Google aims to align the MV3 update with emerging standards for extensions. This unification seeks to reduce fragmentation in the extension ecosystem, allowing developers to create extensions compatible with multiple browsers, such as Chrome, Edge, and Firefox.

Google argues that these changes will enhance user privacy and security and even improve Chrome browser performance, citing the need to protect users from "malicious" extensions that could misuse their access to sensitive data like user credentials and other personal information through means like the \textit{WebRequest} API. Recent improvements to the MV3 update, including increased flexibility in blocking rules, reflect Google’s response to feedback from ad blocker providers \cite{Dunk2023}. However, the concerns regarding a potential reduction in ad-blocking effectiveness underline the ongoing debate between efforts to enhance user privacy and security and the perceived impact on ad blocker effectiveness.

From the perspective of ad blocker providers, the MV3 update introduces restrictions that could reduce the effectiveness of ad blockers even though the changes might improve the security and privacy of other extensions (e.g., coupon code finders, password managers, productivity tools; \cite{Ghostery2021}). They argue, for example, that the new \textit{DeclarativeNetRequest} API changes how ad blockers interact with HTTP requests. Unlike the \textit{WebRequest} API, which allows ad blockers to monitor and block ads and trackers in real-time, the \textit{DeclarativeNetRequest} API requires them to operate based on a predefined set of rules. Additionally, the MV3 update enforces a maximum of 30,000 blocking rules per extension while maintaining a collective restriction of 330,000 predefined blocking rules for all extensions in a browser—the latter raising concerns for ad blocker providers about the reduced effectiveness of a single MV3 ad blocker when used with multiple other MV3 ad blockers simultaneously \cite{Seregin2022}. These restrictions on blocking rules might weaken the flexibility and adaptability of ad blockers in dealing with new or unexpected ads and trackers.

Further, replacing \textit{background pages} with \textit{service workers} to improve browser performance raises practical concerns for ad blocker providers. They argue that short-lived \textit{service workers} can disrupt the continuous monitoring of HTTP requests, which might lead to "ad flickering" (i.e., short re-appearances of website ads) and put ad blockers "[...] into a kind of sleep mode" \cite{Seregin2022}. More precisely, because Chrome automatically suspends an extension’s \textit{service worker} after some time, all global variables, registered listeners and timers are wiped. That means an ad blocker must reload its state from storage and attach listeners immediately on wake‑up or it will miss the very event that triggered the restart.

This loss of state and potential event‑drop can interrupt or delay filtering, so ad blocker providers must redesign their code to persist these challenges \cite{eyeo2024}.  Ad blocker providers also claim that their ad blockers are already highly performant and do not significantly impact browser speed, questioning the value of Google’s change in terms of improving browser performance \cite{Ghostery2021}. \citet{Borgolte2020} addressed that concern and empirically demonstrated that MV2 ad blockers do not significantly impact browser speed.

Thus, while Google’s MV3 update seeks to enhance user privacy, security, and potentially Chrome browser performance, it introduces changes that raise concerns for ad blocker providers. These new restrictions and technical requirements challenge their ability to block ads and trackers effectively. As Google phases out MV2 extensions, ad blocker providers are grappling with innovating and adapting to changes introduced with MV3’s update to maintain their role in providing users with improved privacy and ad-free browsing experiences \cite{Adblock2024, Brinkmann2022, Orlova2023, Meshkov2021, Meshkov2023}.

\subsection{Changes and Workarounds to Google's MV3 Update}

The MV3 update has been argued to reduce ad blocker effectiveness (e.g., \cite{Cronin2019}). This argument forms the basis of potential implications for users, ad blocker providers, Google, and website publishers. However, the assumption that the MV3 update negatively impacts the core functionality of ad blockers remains unverified, and recent developments may suggest otherwise \cite{Dunk2023, Meshkov2023}.

Ad blocker providers have worked extensively to alleviate the potentially negative impact of the MV3 update through workarounds. For instance, AdGuard converted its traditional blocking rules into Chrome’s \textit{DeclarativeNetRequest} format and devised methods to minimize the impact of \textit{service workers} inactivation. Additionally, AdGuard uses a workaround provided by Google to address this issue \cite{Adguard2024}. uBlock Origin introduced uBlock Origin Lite—a fully declarative MV3 ad blocker—that circumvents delays from inactive \textit{service workers} and ensures proper filtering at browser launch \cite{uBlockOrigin2024}. Similarly, Adblock Plus implemented differential filter list updates as a workaround to MV3’s restrictions. The company behind Adblock Plus, eyeo, publicly detailed its approach to handling \textit{service workers} suspensions \cite{Adblock2024, eyeo2024}. Specifically, it employs "fuzz tests" that deliberately suspend \textit{service workers} during testing, ensuring that ad-filtering functionality remains intact even when \textit{service workers} unexpectedly terminate.

Google itself changed the MV3 update to address some of the concerns raised by ad blocker providers. After receiving feedback, Google increased the initially low number of Chrome blocking rules in consultation with ad blocker providers. Additionally, Google introduced improvements such as case-insensitive URL filtering and adjustments to \textit{service workers} lifetimes to alleviate extension performance issues \cite{Dunk2023}.

Despite these changes and workarounds, ad blocker providers continue to warn about potentially reduced core functionality \cite{Adguard2024, uBlockOrigin2024, Adblock2024, Stands2024}. According to them, some problems persist. For example, the inability to define rules based on top-level context \cite{uBlockOrigin2024}, a persistent limit of simultaneously active filter lists \cite{Adblock2024}, or the fact that \textit{service workers} might still be put to sleep, despite workarounds \cite{Adguard2024}. Thus, whether these adaptations have fully alleviated the technical challenges introduced by the MV3 update remains uncertain. The MV3 update could lead to reduced ad blocker effectiveness, no actual impact, or even improvements, though the latter would be unexpected.

If the impact is negative, reducing ad blocker effectiveness could lead to higher visibility of ads for users. It could also increase users’ privacy risks due to greater exposure to online tracking. As a result, users might consider alternatives to Chrome, such as Brave or Vivaldi, which include built-in ad blockers \cite{Ashwin2022}. Additionally, ad blocker providers might face higher development costs to maintain both MV3 and MV2 instances of their ad blockers.

If the MV3 update does not affect ad blocker effectiveness—for instance, if ad blocker providers successfully adapted to its restrictions—users could benefit. Their privacy and ad-blocking experience would remain intact while they gained the privacy and security enhancements introduced by MV3. The update could also create a safer browsing experience and mitigate risks from "malicious" MV2 extensions. Ad blocker providers may not need to maintain separate MV3 and MV2 instances of their software.

The unexpected scenario of a positive impact, where ad blocker providers adapt to MV3’s restrictions and even innovate to improve effectiveness, could lead to more effective ad blocking, enhancing user privacy and creating an even more ad-free browsing experience. Users would encounter fewer ads and fewer trackers. Ad blocker providers would not face increased development costs while demonstrating strong adaptability and innovation. In contrast, Google and website publishers would experience a decline in ad impressions, potentially sacrificing some ad revenue while reinforcing user trust.

\subsection{Reactions of Other Browser Vendors}

Many other browser vendors have approached the MV3 update with caution. Several leading ones such as Brave, Opera, Vivaldi, Firefox, and Safari accepted MV3 extensions while still allowing at least some extensions to run under MV2. For instance, Brave migrated to MV3 yet continues to expose AdGuard, uBlock Origin, uMatrix, and NoScript extensions through its own backend, reflecting a more supportive stance than in its initial roadmap \cite{Brave2025}. Vivaldi, by contrast, states it is effectively obliged to phase out MV2 support, but emphasizes that its built‑in anti‑tracking and ad‑blocking features should offset any major concerns regarding the MV3 update \cite{Vivaldi2024}.

Safari \cite{AppleDevSafariWebExtensions2022} and Firefox \cite{MozillaMV32025}, meanwhile, promise full ongoing support for both MV2 and MV3 extensions, and have announced no retirement date for MV2. On the other end of the spectrum, Microsoft Edge has been least skeptical from the outset: since July 2022 it has blocked new MV2‑based submissions of extensions, arguing that this reduces fragmentation for developers \cite{EdgeMV32020}, mirroring Google's argumentation for the MV3 update.

In summary, because Chrome commands the dominant market share, many other browser vendors appear to have aligned themselves with Google’s new update, willingly or unwillingly.

\section{Related Work}\label{sec:related-work}

In this section, we summarize key findings from related work on ad blocker effectiveness, focusing on their ad-blocking and anti-tracking effectiveness. For an overview of related work, see Table~\ref{tab:lit-review-table} in Appendix~\ref{appendix:lit-review}.

\citet{Wills2016} explored the effectiveness of MV2 ad blockers by comparing and showing variations in said effectiveness. For example, uBlock’s default settings showed robust anti-tracking effectiveness, while other ad blockers like Blur and Disconnect provided limited anti-tracking effectiveness. Ghostery, Adblock Plus, and AdGuard required manual settings configuration for significant anti-tracking effectiveness.

\citet{Garimella2017} expanded on \citet{Wills2016}, highlighting uBlock’s superior ad-blocking and anti-tracking effectiveness, as compared to other ad blockers. Their study showed a paradox of using ad blockers for user privacy: some can even add extra trackers on websites users visit, thereby harming it instead of improving it.

\citet{Merzdovnik2017} and \citet{Borgolte2020} explored the anti-tracking effectiveness of ad blockers, with the former observing challenges ad blockers face in blocking less-known trackers and fingerprinting scripts. The latter highlighted the broader benefits of ad blockers in improving user privacy and browsing experiences.

Additionally, other studies contributed to the existing knowledge in this area (e.g., \cite{Balebako2012, Englehardt2016, Frik2020, Gervais2017, Ikram2017, Malloy2016, Mayer2012, Pantelaios2024}); thus, while above studies provide valuable insights, this list is certainly not exhaustive. Still, it offers a comprehensive overview of related work with a similar browser-based experimental setup, focusing on measuring the MV2 ad blocker effectiveness. This work differs from those studies in comparing the effectiveness of MV3 to MV2 ad blockers.

Other related works share a similar browser-based experimental setup but have different aims. For example, \citet{Demir2024} researched the effectiveness of cookie banner interaction tools and their impact on users’ privacy, while \citet{Datta2015} researched how users’ browsing history (i.e., user behavior and choice) impacts the Google text ads shown to them. This work differentiates from those studies in its aim.

\section{Measurement Setup}\label{sec:measurement-setup}

\subsection{Selection of MV3 and MV2 Ad Blockers}\label{sec:selection-of-ad-blockers}

To achieve the study’s aim and answer the research questions, we identified ad blockers available in both MV3 and MV2 instances from the same ad blocker provider. This approach enabled a direct comparison between the two ad blockers, as they come from the same ad blocker provider but differ in their technical foundations due to the manifest version.

To identify suitable ad blockers for the study, we conducted a search on the Google CWS\footnote{\href{https://chromewebstore.google.com}{https://chromewebstore.google.com}} at the beginning of this study (December 2023), using the key phrases "manifest version 3 (MV3) ad blocker" and "ad blocker MV3." From the pool of results, some ad blockers advertised widespread applications across all websites; others indicated that they functioned exclusively on YouTube, while a few integrated ad blocking as an additional feature to their core functionality (e.g., a VPN). For this study, we selected those ad blockers advertising widespread applications across all websites that offered ad blocking as their core functionality.

We further refined the selection based on the following criteria:

\begin{itemize}
\item \textbf{Early MV3 Adoption}: The ad blockers chosen for this study were among the first to adapt to MV3. This early adoption demonstrates a proactive approach and commitment to staying current with the evolving ad-blocking landscape.
\item \textbf{Substantial User Base}: Each ad blocker selected for this study has a substantial user base, according to CWS statistics. Specifically, we ensured that the MV2 instance of each selected ad blocker had at least 1 million users. This criterion guarantees that the findings of our study are applicable to a broad audience and focus on widely adopted ad blockers.
\item \textbf{Diverse Set of Ad Blocker Providers}: Our selection represents a diverse set of ad blocker providers, each with its unique approach to ad blocking. This diversity allows for a more comprehensive comparison and analysis of their MV3 and MV2 ad blockers.
\end{itemize}

This process unveiled four ad blocker providers: Adblock Plus, with "Adblock Plus - free ad blocker" (MV2) and "Adblock Plus - free ad blocker" (MV3); AdGuard, with "AdGuard AdBlocker" and "AdGuard AdBlocker (MV3 Beta)"; Stands, with "Stands AdBlocker" and "Fair AdBlocker MV3 (Beta)"; and uBlock, with "uBlock Origin" and "uBlock Origin Lite."

Table~\ref{tab:table-2} provides an overview of the MV3 and MV2 ad blockers used in the browser-based experiment, highlighting their versions, last update dates, and the number of Chrome users according to CWS as of August 16, 2024.

\begin{table*}[!ht]
    \centering
    \caption{Overview of manifest version 3 (MV3) and manifest version 2 (MV2) ad blockers used in the browser-based experiment. We report the "Number of Chrome Users" for each ad blocker from the Chrome Web Store as of August 16, 2024.}
    \label{tab:table-2}
    \begin{tabular}{ccrcrrr}
    \toprule
        \parbox{1.8cm}{\centering \textbf{Ad} \\ \textbf{Blocker} \\ \textbf{Provider}} & \textbf{\#} & \textbf{Ad Blocker} & \parbox{1cm}{\centering \textbf{Manifest} \\ \textbf{Version} \\ \textbf{(MV)}} & \parbox{1cm}{\textbf{Version}} & \textbf{Last Update Date} & \parbox{1.8cm}{\centering \textbf{Number of} \\ \textbf{Chrome} \\ \textbf{Users}} \\
    \midrule
         \multirow{2}{1.8cm}{Adblock Plus} & 1 & Adblock Plus - free ad blocker & 2 & \texttt{3.25.1} & April 3, 2024 & 45,000,000+ \\
          & 2 & Adblock Plus - free ad blocker & 3 & \texttt{4.5.1} & August 14, 2024 & 40,000,000+ \\
    \midrule
         \multirow{2}{1.8cm}{AdGuard} & 3 & AdGuard AdBlocker & 2 & \texttt{4.3.53} & May 30, 2024 & 13,000,000+ \\
          & 4 & AdGuard AdBlocker (MV3 Beta) & 3 & \texttt{5.0.33} & August 12, 2024 & 40,000+ \\
    \midrule
         \multirow{2}{1.8cm}{Stands} & 5 & Stands AdBlocker & 2 & \texttt{2.1.24} & August 13, 2024 & 1,000,000+ \\
          & 6 & Fair AdBlocker MV3 (Beta) & 3 & \texttt{2.1.10} & April 18, 2023 & 1,000+ \\
    \midrule
         \multirow{2}{1.8cm}{uBlock} & 7 & uBlock Origin & 2 & \texttt{1.59.0} & August 5, 2024 & 34,000,000+ \\
          & 8 & uBlock Origin Lite & 3 & \texttt{2024.8.12.902} & August 13, 2024 & 300,000+ \\
    \bottomrule
    \end{tabular}
\end{table*}

We used the current versions of each ad blocker available as of August 16, 2024. According to CWS statistics, more than 133 million users actively use these eight ad blockers worldwide, with significantly fewer users currently using their MV3 instances—which reflects the study’s relevance to a substantial user base. Moreover, the anticipation of a user base transition from MV2 to MV3 instances, as Google plans to fully phase out MV2 extensions by July 2025, further emphasizes the timeliness of this study in evaluating ad blocker effectiveness.

\subsection{Description of Browser Experiment and Data Collection Process}\label{sec:browser-experiment}

The central idea behind the browser-based experiment is to simulate the following scenario: a group of four users, each with a different approach to online privacy, visits the same website at about the same time. One user visits it without any ad blocker, serving as a baseline, while the other two employ MV3 and MV2 ad blockers, respectively; the last user uses multiple MV3 ad blockers simultaneously.

\begin{figure*}[ht]
  \centering
  \includegraphics[width=\linewidth]{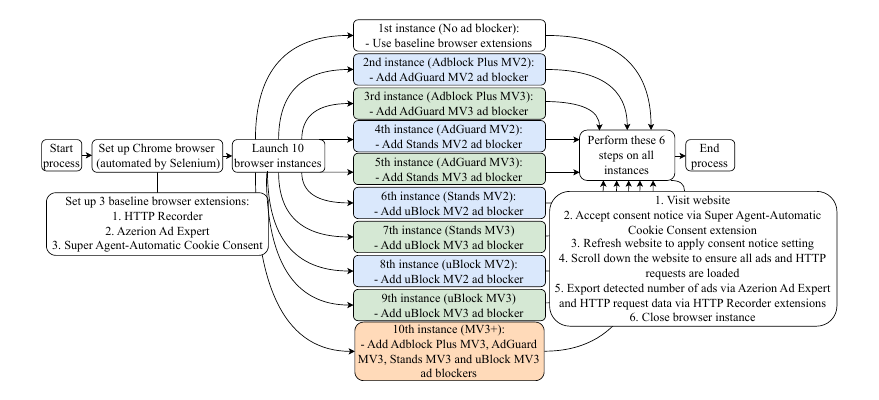}
  \caption{Illustration of the browser experiment and data collection process. We abbreviate manifest version 2 and manifest version 3 to "MV2" and "MV3", respectively. The blue-shaded cells represent the MV2 ad blocker group, which includes Adblock Plus MV2, AdGuard MV2, Stands MV2 and uBlock MV2 ad blockers. The green-shaded cells represent the MV3 ad blocker group, which includes Adblock Plus MV3, AdGuard MV3, Stands MV3 and uBlock MV3 ad blockers. The orange-shaded cell represents the MV3+ ad blocker group, which includes the four MV3 ad blockers (Adblock Plus MV3, AdGuard MV3, Stands MV3 and uBlock MV3 ad blockers).}
  \label{fig:figure-2}
\end{figure*}

We conducted a browser-based experiment to translate that scenario into a controlled experimental setup (Figure \ref{fig:figure-2}). We ran all crawls on a local machine using the Google Chrome browser, reflecting everyday user practice, with a European residential IP address. Using a local machine kept the setup simple, avoided VPN or proxy artifacts, and held geography constant so that any differences reflect MV3 vs. MV2 rather than location. Because all measurements originated from a European residential IP, and GDPR typically reduces third-party activity (e.g., \cite{Peukert2022, Johnson2023, Miller2025}) due to compliance risks irrespective of cookie consent, EU-origin visits tend to activate fewer third parties. As a result, the absolute ad and tracker counts we report are likely conservative relative to a non-EU vantage point. We automated the experiment using Selenium (Python) and a Chrome \textit{user-agent} so that webpages would, in most cases, serve content as they would for a standard visitor.

We ran ten separate browser instances in parallel, each one representing a unique user—one without an ad blocker and the others with a selection of ad blockers from Adblock Plus, AdGuard, Stands, and uBlock, including instances of MV2, MV3, and a combination of these MV3 ad blockers (MV3+). This parallel setup was crucial to maintaining uniform testing conditions and to measure each ad blocker’s ad-blocking and anti-tracking effectiveness under identical web browsing conditions. Running the instances in parallel ensures that all ad blockers are tested under the same conditions simultaneously.

Three extensions were essential for thorough data collection in each browser instance:

\begin{itemize}

    \item \textbf{HTTP Recorder}: Developed for this study, the "HTTP Recorder" extension captured HTTP requests upon a webpage visit. This custom tool uses APIs like \textit{WebRequest}, recording detailed traffic data upon a webpage visit necessary for analyzing its network activity to determine the number of blocked trackers. Given the restrictions set by the CWS’s review process, such a level of detail is beyond what standard extensions provide, necessitating the development of such an extension.

    \item \textbf{Azerion Ad Expert}\footnote{\href{https://chromewebstore.google.com/detail/azerion-ad-expert/nndadbimjipilgfojofhpjjkhgflkihc}{https://chromewebstore.google.com/detail/azerion-ad-expert/nndadbimjipilgfojofhpjjkhgflkihc}}: This extension was crucial for ad detection upon a webpage visit. It captured the Prebid framework (i.e., \texttt{prebid.js}) and Google Ad Manager (GAM) auction details for display and video ads and offered insights into header bidding auctions. Its functionality to record ad slots on the webpage was crucial for verifying the ad-blocking process and detecting the number of webpage ads upon a visit.

    \item \textbf{Super Agent–Automatic Cookie Consent}\footnote{\href{https://chromewebstore.google.com/detail/superagent-automatic-cook/neooppigbkahgfdhbpbhcccgpimeaafi}{https://chromewebstore.google.com/detail/superagent-automatic-cook/neooppigbkahgfdhbpbhcccgpimeaafi}}: This extension automated the acceptance of cookie consent forms across websites and ensured ads were loaded as they would be for users who consent to data collection and advertising upon a webpage visit. We configured this extension to accept cookies by default, standardizing the browsing conditions across webpage visits.
    
\end{itemize}

To conduct the browser-based experiment, each browser instance equipped with the above three extensions performed a series of steps in parallel. These steps included visiting a webpage to initiate the browsing session, using "Super Agent–Automatic Cookie Consent" to automatically detect and accept cookie notices to ensure that ads would load upon visiting the webpage, refreshing the webpage to ensure that the cookie consents were applied, and accurately capturing subsequent ad and HTTP request loadings using "Azerion Ad Expert" and "HTTP Recorder", respectively. Additionally, the browser would scroll through the webpage to mimic the engagement level of an average user reading through the homepage of a website and trigger additional ads or HTTP requests. The browser would then export ad and HTTP request data, as these metrics were essential for evaluating the ad blocker effectiveness. Finally, the browser would reset its state for each new website visit to ensure a clean and consistent starting point for each test.

In addition to automated data collection, we captured screenshots and HTML files during each website visit. These additional data sources provide further validation of ad blocker effectiveness and allow for manual verification of detected ads and trackers.

The "stateless" crawling approach emulated the sequential actions of users upon a webpage visit and ensured the experiment’s integrity by preventing data carry-over effects. By resetting the state after each website visit (i.e., a "stateless" crawling approach), the experimental setup guaranteed a uniform baseline for data collection, which is crucial for a valid comparative analysis.

Figure \ref{fig:figure-2} visualizes this experimental setup and the sequence of steps. It clarifies the parallel operation of the browser instances and the approach we used to measure the effectiveness of the ad blockers.

We launched the browser-based experiment on May 15, 2025. To account for webpage dynamism (e.g., \cite{Demir2023}), we conducted five separate measurement runs (runs). Unless stated otherwise, all main sample analyses use per-website averages computed across these five runs for each browser instance; all statistical tests and error bars are based on these per-website averages. A descriptive analysis revealed low variance between runs (see Appendix~\ref{appendix:variance-analysis}). Each run took about 64 hours on average. The total duration of the experiment for our main sample was around 320 hours and 11 minutes, driven by the extensive wait times and webpage scrolling needed for the consistent measurements.

To ensure the cross-browser applicability of our findings, we also launched the browser-based experiment using the Firefox browser. However, conducting the browser-based experiment using the Firefox browser presents a key challenge: unlike in Chrome browser, we could not use the same set of MV3 and MV2 ad blockers. Firefox has announced that it will continue supporting both MV3 and MV2 extensions in the future \cite{Sullivan2024} and currently has no plans to phase out MV2 extensions. While this strategy ensures ongoing availability of MV2 ad blockers in the Firefox browser, it has also delayed the full adaption of ad blocker providers to MV3. As a result, only uBlock Origin provides both MV3 and MV2 instances in the Firefox browser. Complicating matters further, as of late 2024, the Firefox Add-ons Store\footnote{\href{https://addons.mozilla.org/en-US/firefox/}{https://addons.mozilla.org/en-US/firefox/}} removed uBlock Origin Lite—uBlock’s MV3 instance—due to disagreements with its developer \cite{NoahDurham2024}. Although no longer listed in the Firefox Add-ons Store, uBlock Origin Lite remains available for manual installation in the Firefox browser. Despite these challenges, we conducted a limited version of our browser-based experiment using the Firefox browser to ensure the cross-browser applicability of our findings (see Section~\ref{sec:insights-from-robustness-tests} and Appendix~\ref{appendix:firefox}). In this Firefox browser-based experiment, we evaluated uBlock MV3 and MV2 on a main sample of 1,000 websites across five measurement runs beginning July 12, 2025, and obtained consistent results for 824 websites (i.e., websites with valid data across all five runs; others were excluded due to redirects, anti-automation/ad-blocking, or failed instrumentation).

\subsection{Selection of Websites for the Browser-Based Experiment}\label{sec:selection-of-websites}

For the main sample of our browser-based experiment, we selected ad-supported websites to observe the impact of ad blockers and evaluate their effectiveness. For this purpose, we utilized a curated list from BuiltWith\footnote{\href{https://builtwith.com}{https://builtwith.com}}, which comprised data of websites relying on the Prebid framework. From the list of websites provided by BuiltWith on January 9, 2024, we selected the top 1,000 most popular websites according to their Tranco popularity rank \cite{LePochat2019}. The Tranco popularity rank of these websites varied from 17 to 17,590. We selected these websites for their reliance on ad-supported content and their use of the Prebid framework, guaranteeing compatibility with the Azerion Ad Expert extension (see Section~\ref{sec:browser-experiment}). This compatibility is crucial for ensuring the extension accurately records the number of ads, especially for the baseline browser instance using no ad blocker.

Across five separate measurement runs, we consistently obtained results for 924 of these websites, excluding those that redirected users, deployed ad-blocking prevention or anti-automation mechanisms, or were inaccessible. The final list for our main sample represents 924 popular, ad-supported websites across 23 countries.

While our main sample targets popular, ad-supported websites, ad blockers may behave differently on less popular websites or websites with limited resources—for example, because extensions sometimes need to drop less frequently used blocking rules when they hit internal limits on total rule counts. To address this point and improve the external validity of our results, we conducted the browser‑based experiment on two additional stratified samples for robustness tests. Using the same BuiltWith dataset, we stratified:

\begin{itemize}
  \item One sample by website employee count, yielding 191 websites (including many with zero recorded employees), and
  \item Another by Tranco popularity rank, yielding 185 websites (extending as far down as ranks in the millions).
\end{itemize}

These stratified samples therefore cover both ends of the popularity and resource‑availability spectrum, ensuring our findings are not driven solely by high‑traffic, well‑resourced websites. The resulting diversity in organizational size and popularity enhances the generalizability of our conclusions (see Section~\ref{sec:insights-from-robustness-tests} and Appendix~\ref{appendix:stratified-samples}).

\subsection{Definition of Two Metrics for Ad Blocker Effectiveness}\label{sec:definition-of-two-metrics}

We measured ad blockers’ ad-blocking and anti-tracking effectiveness using the number of blocked ads and the number of blocked trackers. To establish a baseline, we measured the number of ads and trackers on a website without ad blockers.

For the first metric, the number of blocked ads, we used the extension Azerion Ad Expert. As mentioned (see Section~\ref{sec:browser-experiment}), this tool recorded the number of display ad slots on a website. By subtracting the measured number of ads on the website while using the ad blocker from the baseline number of ads on the website without using the ad blocker, we obtained the number of blocked ads.

To validate the accuracy of these ad counts, we manually counted ads in collected screenshots from a random subsample of 100 website-ad blocker observations in one of our early runs. A Pearson correlation analysis showed a strong positive correlation (\( r = 0.90 \), 95\% CI \([0.86, 0.93]\), \( p < 0.001 \)), indicating high agreement between our manual and the Azerion Ad Expert ad counts. The ad counts differed by more than one ad in only three website-ad blocker observations, likely due to screenshots not always capturing the full webpage. These results show that Azerion Ad Expert accurately counts ads on webpages.

The second metric measures the number of blocked trackers. We measured this metric from collected HTTP request data. We identified the third-party domain (e.g., \texttt{demdex.net}) for each HTTP request (e.g., visiting the \texttt{cnn.com} website creates an HTTP request \texttt{https://dpm.demdex.net/id?d\_visid\_[...]}). Then, we \\ cross-referenced that third-party domain with WhoTracks.me's database\footnote{\href{https://whotracks.me}{https://whotracks.me}} of known trackers (e.g., Adobe Audience Manager tracker uses a \texttt{demdex.net} domain; \cite{Karaj2018}). By subtracting the measured number of trackers on the website while using the ad blocker from the baseline number of trackers on the website without using the ad blocker, we obtained the number of blocked trackers.

\section{Results}\label{sec:results}

\subsection{Descriptive Results for Ad Blocker Effectiveness}\label{sec:descriptive-results}

\begin{figure}[!htb]
    \centering
    \includegraphics[width=\columnwidth]{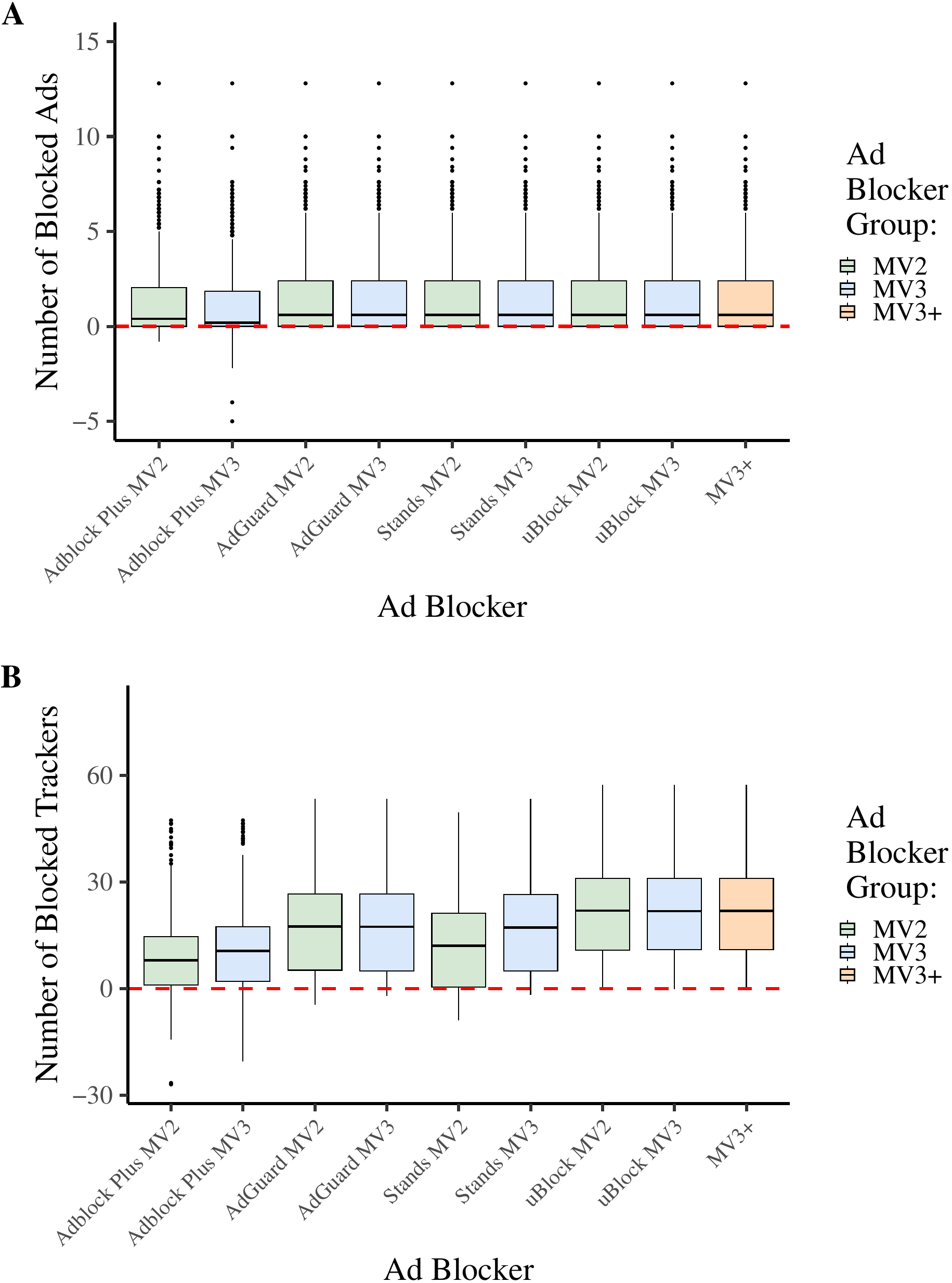}
    \caption{
        Distribution of the number of blocked ads (Panel A) and trackers (Panel B) per website and browser instance. The red dashed horizontal line at 0 highlights the threshold below which ad-blocking or anti-tracking effectiveness is negative. Multiplying the 9 browser instances (Adblock Plus MV2, Adblock Plus MV3, AdGuard MV2, AdGuard MV3, Stands MV2, Stands MV3, uBlock MV2, uBlock MV3, and MV3+) with the number of websites (924) yields the number of observations (N = 8,316). Values shown are per-website averages across five runs.
    \label{fig:figure-3}
    }
\end{figure}

Figure~\ref{fig:figure-3} presents box plots that summarize the number of blocked ads (Panel A) and trackers (Panel B) for MV3 and MV2 ad blockers as well as for the MV3+ ad blocker group.

\textbf{Blocked Ads (Panel A):}
The distributions for the number of blocked ads are similar across ad blockers, indicating comparable ad‐blocking effectiveness. MV2 ad blockers (Adblock Plus MV2, AdGuard MV2, Stands MV2, and uBlock MV2) show medians ranging from about 0.4 to 0.6 blocked ads. Their MV3 counterparts exhibit similar medians—with the exception of Adblock Plus MV3, showing a lower median (around 0.2) and a minimum of –5. We obtain negative values when a website displays more ads with the ad blocker than without it (e.g., 12 ads with the ad blocker versus 10 without the ad blocker yields –2 ads per website and browser instance). All ad blocker groups, including the MV3+ ad blocker group, consistently show an upper outlier at 13 blocked ads, indicating that certain websites repeatedly display a high number of ads.

\textbf{Blocked Trackers (Panel B):}
The anti‐tracking effectiveness varies among ad blockers. Adblock Plus's MV3 and MV2 instances exhibit medians of about 8–10.6 blocked trackers, while AdGuard's MV3 and MV2 instances yield medians of about 17 blocked trackers. In contrast, the uBlock's MV3 and MV2 instances and the MV3+ ad blocker group show higher medians, around 22 blocked trackers. Although most ad blockers display similar central tendencies, some observations fall below 0 blocked trackers—for example, Adblock Plus MV2 reaches a minimum of –27 blocked trackers, meaning that the number of blocked trackers sometimes exceeds the baseline. Notably, negative values also appear for other instances (e.g., down to –20.6 for Adblock Plus MV3 and –1.8 for Stands MV3). Moreover, the MV3+ ad blocker group achieves a higher median anti‐tracking effectiveness than several MV2 ad blockers (Adblock Plus MV2, AdGuard MV2, and Stands MV2) and is similarly higher than most MV3 ad blockers. The anti-tracking effectiveness of uBlock MV3 appears to drive these results in the MV3+ ad blocker group, suggesting that multiple MV3 ad blockers do not hinder each other through the global rule limit.

\subsection{Bi-Group Comparison: MV3 vs. MV2 Ad Blocker Effectiveness}\label{sec:bi-group-comparison}

\begin{figure}[!htb]
  \centering
  \includegraphics[width=\linewidth]{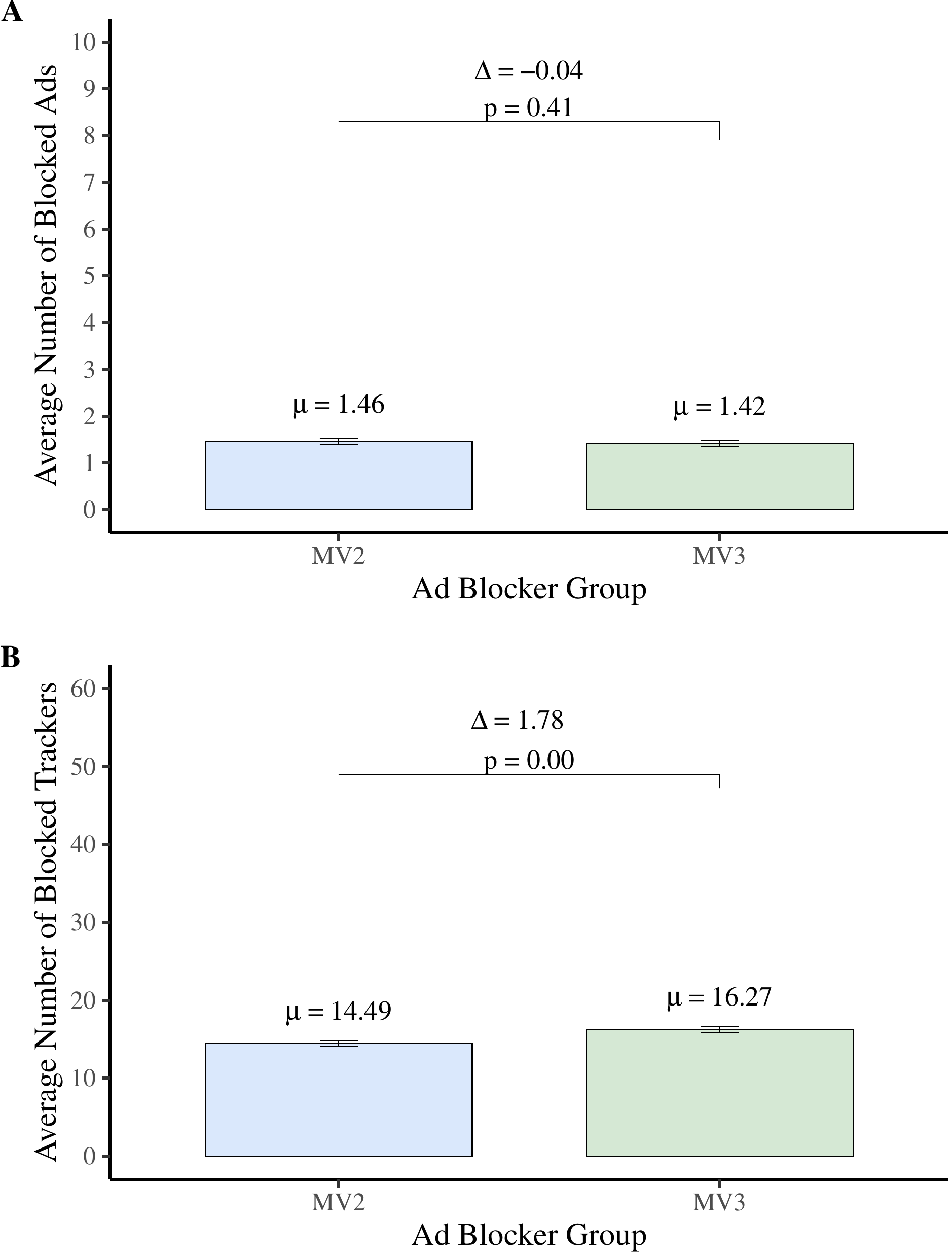}
  \caption{Number of blocked ads and trackers between manifest version 3 (MV3) and manifest version 2 (MV2) ad blocker groups. This figure shows independent t-test comparisons between the mean values of two ad blocker groups, namely MV3 and MV2, regarding the number of blocked ads (Panel A) and trackers (Panel B). The MV3 ad blocker group includes Adblock Plus MV3, AdGuard MV3, Stands MV3, and uBlock MV3 ad blockers, while the MV2 ad blocker group includes Adblock Plus MV2, AdGuard MV2, Stands MV2, and uBlock MV2 ad blockers. We exclude the MV3+ ad blocker group from these comparisons. Multiplying the 8 browser instances (Adblock Plus MV2, Adblock Plus MV3, AdGuard MV2, AdGuard MV3, Stands MV2, Stands MV3, uBlock MV2, uBlock MV3) with the number of websites (924) yields the number of observations (N = 7,392). Values shown are per-website averages across five runs; error bars represent $\pm 1$ SE across websites.
  }
  \label{fig:figure-4}
\end{figure}

Transitioning from descriptive results, we used an independent-sample t-test to compare the effectiveness of the MV3 and MV2 ad blocker groups, as depicted in Figure \ref{fig:figure-4}.

\textbf{Blocked Ads (Panel A):} Panel A of Figure~\ref{fig:figure-4} shows that there is no significant difference in ad-blocking effectiveness between the MV3 and MV2 ad blocker groups ($\Delta = -0.04$, $p = 0.41$). The mean number of blocked ads is about the same for both groups, and the error bars indicate similar variability within each group.

\textbf{Blocked Trackers (Panel B):} Panel B of Figure~\ref{fig:figure-4} illustrates a significant difference in anti-tracking effectiveness between the MV3 and MV2 ad blocker groups ($\Delta = 1.78$, $p < 0.001$). The difference of about 2 blocked trackers in the mean number of blocked trackers is statistically significant, and the variability, as indicated by the error bars, is comparable between the groups.

\subsection{Individual Ad Blocker Comparisons: MV3 vs. MV2 and MV3+ vs. MV3 Ad Blocker Effectiveness }\label{sec:individual-comparisons}

\begin{figure*}[!htb]
    \centering
    \includegraphics[width=\linewidth]{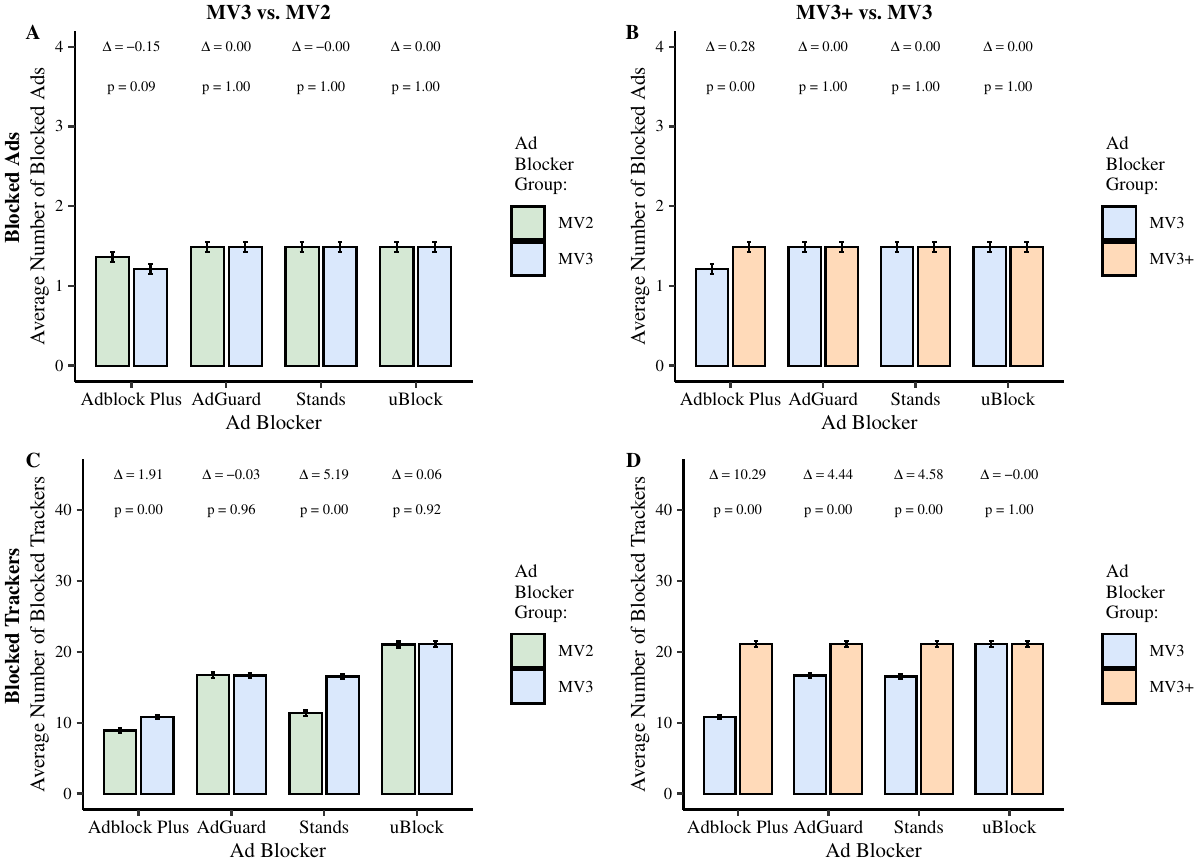}
    \caption{
        Number of blocked ads and trackers between individual ad blockers and the MV3+ ad blocker group. Panels A and C show independent t-test comparisons between the mean values of the MV3 and MV2 ad blockers for the number of blocked ads and trackers, respectively. We exclude the MV3+ ad blocker group from these comparisons. Multiplying the 8 browser instances (Adblock Plus MV2, Adblock Plus MV3, AdGuard MV2, AdGuard MV3, Stands MV2, Stands MV3, uBlock MV2, uBlock MV3) with the number of websites (924) yields the number of observations (N = 7,392). Panels B and D show comparisons between the MV3+ ad blocker group and the individual MV3 ad blockers for the number of blocked ads and trackers, respectively. Multiplying the 5 browser instances (Adblock Plus MV3, AdGuard MV3, Stands MV3, uBlock MV3, and MV3+ ad blocker group) with the number of websites (924) yields the number of observations (N = 4,620). Values shown are per-website averages across five runs; error bars represent $\pm 1$ SE across websites.
    \label{fig:figure-5}
    }
\end{figure*}

Next, we performed a series of independent-sample t-tests to compare the effectiveness of MV3 ad blockers from the same ad blocker providers to their MV2 counterparts, and the MV3+ ad blocker group to the individual MV3 ad blockers (Figure~\ref{fig:figure-5}).

\textbf{Blocked Ads (Panels A and B):}
In Panel A, the comparison between the MV3 and MV2 ad blockers reveals no significant differences in their ad-blocking effectiveness.

In Panel B, comparisons between the MV3+ ad blocker group and the MV3 ad blockers show a significant difference for Adblock Plus ($\Delta=0.28$, $p < 0.01$), while AdGuard, Stands, and uBlock show no significant differences.

\textbf{Blocked Trackers (Panels C and D):}
Panel C shows higher anti-tracking effectiveness for Stands MV3 compared to its MV2 instance ($\Delta=5.19$, $p < 0.001$). The mean number of blocked trackers for Stands MV2 is about 11.3, while for Stands MV3 it is about 16.5. This increase of 5.19 blocked trackers represents a 45.9\% improvement for Stands MV3: $(\frac{16.5-11.3}{11.3})\times 100\% = 45.9\%$. In addition, Adblock Plus MV3 also exceeds its MV2 instance ($\Delta=1.91$, $p < 0.001$), representing a 21.5\% improvement for Adblock Plus MV3: $(\frac{10.8-8.89}{8.89})\times 100\% = 21.5\%$. AdGuard and uBlock show no significant differences.

In Panel D, the comparison between the MV3+ ad blocker group and the MV3 ad blockers reveals higher anti-tracking effectiveness of the MV3+ ad blocker group compared to Adblock Plus MV3 ($\Delta=10.29$, $p < 0.001$), AdGuard MV3 ($\Delta=4.44$, $p < 0.001$), and Stands MV3 ($\Delta=4.58$, $p < 0.001$). These results demonstrate that using multiple MV3 ad blockers simultaneously (MV3+) enhances anti-tracking effectiveness by about 95.3\% over Adblock Plus MV3 [$(\frac{10.29}{10.8})\times 100\% = 95.3\%$], 26.6\% over AdGuard MV3 [$(\frac{4.44}{16.7})\times 100\% = 26.6\%$], and 27.8\% over Stands MV3 [$(\frac{4.58}{16.5})\times 100\% = 27.8\%$].

\subsection{Insights from Robustness Tests}\label{sec:insights-from-robustness-tests}

In this section, we summarize the key findings from five robustness tests we conducted to verify the consistency and reliability of our results. Table~\ref{tab:table-3} summarizes the robustness tests, fundamental concerns they address, and their results. We provide more details in Appendix~\ref{appendix:robustness}.

\begin{table*}[ht]
    \centering
    \caption{Summary of robustness tests.}
    \label{tab:table-3}
    \begin{tabular}{p{4.2cm} p{5cm} p{5.8cm} c}
    \toprule
    \textbf{Robustness Test} & \textbf{Fundamental Concern} & \textbf{Summary of Result} & \textbf{Appendix} \\
    \midrule
    Effectiveness of MV3 Ad Blockers Across Different Samples & Our main sample might not be representative of the overall population of websites; need for additional, more representative samples & Findings are consistent across two stratified samples (stratified by website employee count and Tranco popularity rank), confirming the robustness of our main results. & \ref{appendix:stratified-samples} \\
    \midrule
    Alternative Measures for Ad Blocker Effectiveness & The original measures of ad blocker effectiveness might fail to capture certain types of ads or lesser-known trackers; need for alternative measures. & Results remain consistent using HTML file size (kb) of loaded websites and number of blocked third-party domains, confirming the robustness of our main results. & \ref{appendix:alternative-anti-tracking} \\
    \midrule
    Effectiveness of Early MV3 Ad Blockers & Instances of early MV3 ad blockers might have been less effective; need to assess if their effectiveness changed over time. & Early MV3 instances are as effective as later ones; effectiveness has remained consistent over time. & \ref{appendix:early-mv3-ad-blockers} \\
    \midrule
    Effectiveness of MV3 Ad Blockers on Firefox & Findings might be specific to Chrome; need to verify results across different browsers. & Results on Firefox align with those on Chrome; uBlock's effectiveness remains consistent across browsers. & 
    \ref{appendix:firefox} \\
    \midrule
    Visual Inspection of MV3 and MV2 Ad Blocker Screenshots & Potential ad flickering, unintended blockage of legitimate content or cosmetic differences might not be fully captured by our metrics. & We detected no noticeable ad flickering or website breakage in screenshots; we observed some differences in the visibility of cosmetic placeholders. & 
     \ref{appendix:visual_inspection} \\
    \bottomrule
    \end{tabular}
\end{table*}

The first robustness test (see Appendix~\ref{appendix:stratified-samples}) examined whether our findings hold despite variations in sample size and website selection. We repeated our analysis on two separate stratified samples. These samples were stratified by website employee count, since websites with varying resources might serve ads differently, and Tranco popularity rank, creating samples more representative of the overall population of websites in resources and popularity. In this test, the Tranco popularity rank of tested websites varied from 158 to more than one million, and the number of employees from 0 to 10,000, covering a wide range of website types. Specifically, in the sample stratified by website employee count (191 websites), MV3 and MV2 did not differ in blocked ads; for trackers, only Stands improved under MV3 ($\Delta = 7.03$, $p < 0.001$). In the sample stratified by Tranco popularity rank (185 websites), MV3 and MV2 did not differ in either metric. In both stratified samples, the MV3+ group blocked significantly more trackers than Adblock Plus MV3, AdGuard MV3, and Stands MV3 (no significant difference vs. uBlock MV3). Taken together, these stratified-sample results are largely consistent with those of our main sample, reinforcing the robustness of our main findings.

The second robustness test (see Appendix~\ref{appendix:alternative-anti-tracking}) used alternative measures of ad blocker effectiveness, namely HTML file size of loaded websites and the number of blocked third-party domains. We measured the HTML file size (kb) of loaded websites as an alternative measure for ad-blocking and anti-tracking effectiveness of ad blockers. T-tests between the MV3 and MV2 instances of ad blockers increase confidence in our main results by showing mostly non-significant differences between manifest versions. An exception is the Stands MV3 ad blocker, which caused significantly lower HTML file sizes than its MV2 counterpart, in line with previous findings. Similarly, the number of blocked third-party domains is an alternative measure for anti-tracking effectiveness to the number of blocked trackers. While the number of blocked trackers is a measure derived from third-party domains cross-referenced with the WhoTracks.me database, sensitive to well-known trackers, the number of blocked third-party domains counts all distinct third-party domains blocked, offering a broader view of the ad blockers’ anti-tracking effectiveness. Despite possible variations due to this more granular measure, the pattern mirrored our tracker results at the individual level: Adblock Plus and Stands blocked more domains under MV3, while AdGuard and uBlock showed parity between MV3 and MV2. Overall, these alternative measures reinforce our main findings.

The third robustness test (see Appendix~\ref{appendix:early-mv3-ad-blockers}) compared the first versions of MV3 ad blockers (AdGuard MV3 version \texttt{0.3.9} and uBlock MV3 version \texttt{0.1.22.806}) with their MV2 counterparts. This comparison aimed to assess changes in their effectiveness over time, exploring whether ad blocker providers adapted to MV3’s restrictions. Interestingly, the early MV3 versions were equally effective as their later versions, suggesting that the effectiveness of MV3 ad blockers has remained mostly consistent over time. This finding suggests that when ad blocker providers first introduced their MV3 instances of ad blockers, they were likely confident in their effectiveness. The core functionality of these two ad blockers did not fluctuate significantly across subsequent releases.

The fourth robustness test (see Appendix~\ref{appendix:firefox}) evaluated ad blocker effectiveness on the Firefox browser to address concerns that our Chrome results might be browser-specific. We conducted this robustness test with uBlock as it is currently the only ad blocker on Firefox with a MV3 and MV2 instance. As in our browser-based experiment, we developed a custom HTTP Recorder extension to count the number of blocked trackers, and counted the HTML file size of loaded websites—as Azerion Ad Expert is not available for Firefox browser. We found that our results on the Firefox browser align with those on Chrome browser, indicating that uBlock's effectiveness remained consistent across browsers and manifest versions. This result indicates that our findings are applicable across the two browsers.

Lastly, we conducted a fifth robustness test by visually inspecting the captured screenshots during the crawling process. These screenshots were taken as the browser scrolled through each webpage, allowing us to compare a subset of the tested MV3 and corresponding MV2 browser instances (100 each) side by side. Our goal was to detect additional ad flickering (the visibility of ads, that were not present in our ad counts), unintended blockage of legitimate content (content that is considered regular website functionality irrespective of ads), or cosmetic placeholders (the visibility of empty placeholders on webpages). Two evaluators independently evaluated outcomes based on a provided set of questions, one per outcome. We report results as Evaluator 1 (Evaluator 2). The two evaluators found no substantial ad-flickering discrepancies between MV3 and MV2: 4 (4) missed ads, split evenly across MV3 and MV2. They also observed no major breakage differences: 5 (5) MV3-only versus 5 (3) MV2-only cases. However, they found that MV3 left over cosmetic placeholders in 21\% (22\%) of cases versus 0\% (0\%) under MV2. Further details on how we conducted this test and examples can be found in Appendix~\ref{appendix:visual_inspection}. In summary, we conclude that the MV3 update did not introduce notable additional ad flickering or website breakage, though MV3 ad blockers occasionally produced a less visually appealing experience than their MV2 counterparts due to the increased visibility of cosmetic placeholders.

\section{Limitations and Ethics}\label{sec:limitations-and-ethics}

This study comes with several limitations. First, as in previous studies (e.g., \cite{Demir2024}), this study faces limitations related to the dynamic behavior of websites and extensions. Specifically, it cannot verify whether websites consistently respect a user's choice as dictated by extensions (e.g., Super Agent–Automatic Cookie Consent always accepting cookies across websites) or if an extension (e.g., an ad blocker) could malfunction, impacting the results. While we conducted preliminary testing and validation checks to ensure the browser-based experiment's reliability, the possibility exists that website compatibility fluctuates. Although this would not affect the validity of our results, as these observations would be included as zero-counts in our data, future research efforts could focus on developing methods to verify the extensions' actions and the websites' responses more accurately. This would address whether websites consistently respect these automated choices and further ensure the reliable functioning of the used extensions in browser-based experiments.

Second, consistent with previous research \cite{Demir2024}, the automated browser-based experiment does not replicate all aspects of user interactions, such as visiting sub-pages of a website’s homepage, which introduce different trackers than the website’s homepage \cite{Aqeel2020, Demir2022, Urban2020}. We used the homepage as a uniform entry point for scale and comparability; most ad/analytics tags are installed site-wide, so we expect relative MV3 vs. MV2 differences to be similar across webpage types, even if some sub-pages may show higher numbers of ads or trackers. Future work could explore a broader range of behaviors, including visiting sub-pages.

Third, the study explores ad blocker effectiveness using a European IP address and default ad blocker settings on a select sample of websites, focusing on display ads. Specifically, we examined four popular ad blockers, which do not represent the full ecosystem of ad blockers. Further, we focused on a select number of websites, which are likely skewed towards more popular ad vendors. As a result, the web’s long tail is under-represented. Consequently, we cannot rule out that MV3’s 30,000-rule limit could disproportionately affect less popular websites that rely on rarely used rules. Fully testing the impact of this limit would require a larger set of less popular websites. We also primarily investigated header bidding ads served via Prebid framework, potentially missing other types of served ads (e.g., waterfall-style bidding). Because EU traffic often involves GDPR-driven website changes that reduce third-party activity irrespective of consent, baseline ad and tracker counts are lower. Thus, our absolute counts are conservative (lower-bound) estimates of ad blocker effectiveness. Future work could vary geography, ad blocker settings, and ad types (including video) to broaden external validity.

Fourth, our study cannot exclude that any future changes to the Chrome extension ecosystem might hinder ad blocker effectiveness, as feared by some ad blocker providers \cite{Adguard2024}, or that ad blocker providers get crowded out in the long term (e.g., through increased development and maintenance costs). Future examination of ad blocker effectiveness might therefore reveal nuanced results.

Our study has no conflicts of interest; we are not affiliated with Google (Alphabet) or any ad blocker provider, and our research is independently funded. The browser-based experiment generated automated website visits, which have affected ad impressions. However, the impact is negligible given the limited number of website visits and the scale of online advertising.

\section{Conclusion}\label{sec:conclusion}

This study evaluated the impact of the MV3 update on ad blocker effectiveness, focusing on their ad-blocking and anti-tracking effectiveness. The study relied on a browser-based experiment that tested four popular ad blockers across several website samples and two browsers. The key findings of this study are:

\begin{itemize}

\item \textbf{RQ1: Effectiveness Among MV3 and MV2 Ad Blocker Groups.} The study revealed no difference in the number of blocked ads between MV3 and MV2 groups of ad blockers. For trackers, the MV3 ad blocker group blocked about 1.8 more trackers on average per website than the MV2 group, indicating that the MV3 update did not reduce anti‑tracking effectiveness.

\item \textbf{RQ2: Variability in Effectiveness Among Individual Ad Blockers.} While the ad-blocking effectiveness remained consistent between MV3 and MV2 instances for most ad blockers, Adblock Plus MV3 blocked about 1.9 more trackers (21.5\%) than its MV2 counterpart, and Stands MV3 blocked about 5.2 more trackers (45.9\%) than its MV2 counterpart. This result suggests that these ad blocker providers may have optimized their ad blockers to MV3 or that MV3 allows for enhanced anti‑tracking effectiveness.

\item \textbf{RQ3: Enhanced Anti-Tracking Effectiveness with Multiple MV3 Ad Blockers.} Contrary to the initial concerns of some ad blocker providers, using multiple MV3 ad blockers simultaneously did not reduce their ad‑blocking effectiveness. Instead, it significantly enhanced their anti‑tracking effectiveness. Specifically, the combination of multiple MV3 ad blockers (Adblock Plus MV3, AdGuard MV3, Stands MV3, and uBlock MV3) blocked about 10.3 more trackers (95.3\%) than Adblock Plus MV3 alone, 4.4 more (26.6\%) than AdGuard MV3 alone, and 4.6 more (27.8\%) than Stands MV3 alone, with no difference relative to uBlock MV3. The improvement in anti-tracking effectiveness is driven by the inclusion of the uBlock MV3 ad blocker in the combination, and it is likely due to uBlock's stronger anti-tracking default configurations after installation—as highlighted in related work \cite{Merzdovnik2017, Wills2016}. This finding indicates that concerns regarding the combined use of MV3 ad blockers were unfounded. Notably, this finding represents increased privacy and security for a privacy-aware user who uses multiple MV3 ad blockers simultaneously.

\end{itemize}

In conclusion, this study contributes to understanding the impact of the MV3 update on ad blocker effectiveness. The empirical findings indicate no statistically significant reduction in ad-blocking or anti-tracking effectiveness for MV3 ad blockers compared to their MV2 counterparts. Some of our results even suggest slight improvements in blocking trackers for specific ad blockers following the MV3 update. These slight improvements might stem from filtering out unnecessary blocking rules (e.g. \cite{Snyder2020, Bottger2025}) or continued improvement of ad blockers by their providers.

Our results hold across different website samples, alternative effectiveness metrics, and over time. Moreover, cross-browser experiments yield comparable outcomes, and visual inspection confirms that MV3 ad blockers work effectively without significant ad flickering or loss of functionality. Still, MV3 ad blockers tend to offer a slightly less visually appealing browsing experience than their MV2 counterparts, mainly due to the increased visibility of cosmetic placeholders. In total, these findings are reassuring for users who rely on ad blockers for a more private and ad-free browsing experience.

\section*{Acknowledgements}
\footnotesize

This project has received funding from the European Research Council (ERC) under the European Union’s Horizon 2020 research and innovation program (grant agreement No. 833714). Lazaros Papadopoulos acknowledges financial support from the German Academic Exchange Service (DAAD) and also German Research Foundation (DFG) under the TRR266 program (Project-ID 403041268).
\normalsize

\bibliography{references}


\clearpage

\appendix

\section{Supplementary Material}

\subsection{Overview of Related Work on Ad Blocker Effectiveness}\label{appendix:lit-review}

Table~\ref{tab:lit-review-table} summarizes related work on ad blocker effectiveness, highlighting the measures for ad-blocking and anti-tracking effectiveness, analyzed manifest version(s), and key findings.

\begin{table*}[h!]
  \centering
  \caption{Overview of related work on ad blocker effectiveness. We arrange the table based on the year of publication. We abbreviate manifest version 2 to "MV2" and manifest version 3 to "MV3".}
  \begin{tabular}{cccccc}
    \toprule
    \multirow{2}{2cm}{\centering \textbf{Author}}  &  \multicolumn{2}{c}{\textbf{Measure(s) of Ad Blocker Effectiveness}} & \multicolumn{2}{c}{\parbox{2.5cm}{\centering \textbf{Ad Blocker}\\\textbf{Version(s) Used}\\\textbf{for Comparisons}}} &  \multirow{2}{6cm}{\parbox{6cm}{\textbf{Key Finding(s) on Ad Blocker Effectiveness}}} \\
    \cmidrule(lr){2-5} 
    & \parbox{2cm}{\centering \textbf{Ad-Blocking} \\ \textbf{Effectiveness}} & \parbox{2.5cm}{\centering \textbf{Anti-Tracking} \\ \textbf{Effectiveness}} & \textbf{MV2} & \textbf{MV3} \\
    \midrule
    \parbox{2.5cm}{\centering Wills and \\ Uzunoglu (2016)} & \xmark & \parbox{2.5cm}{Number of blocked\\unique domains in\\HTTP requests} & $\checkmark$ & \xmark & \parbox{6cm}{Different ad blockers have varying levels of effectiveness. By default, uBlock has the best anti-tracking effectiveness, while Blur and Disconnect offer limited protection against trackers. Instead, Ghostery, Adblock Plus, and AdGuard require manual configuration to provide significant anti-tracking effectiveness.} \\
    \midrule
    \parbox{2cm}{\centering Garimella \\ et al. (2017)} & \parbox{2.5cm}{Total size of downloaded and uploaded data; cumulative time for loading the website; wall-clock time for loading the website} & \parbox{2.5cm}{Number of blocked HTTP requests; number of blocked unique domains in HTTP requests} & $\checkmark$  & \xmark  & \parbox{6cm}{uBlock has improved ad-blocking and anti-tracking effectiveness while preserving user privacy compared to other analyzed ad blockers. However, using ad blockers may not always make websites load faster because ad blockers can add extra trackers to the website, harming users' privacy.} \\
    \midrule
    \parbox{2cm}{\centering Merzdovnik \\ et al. (2017)} & \xmark & \parbox{2.5cm}{Number of blocked HTTP requests; number of blocked unique domains in HTTP requests} & $\checkmark$ & \xmark & \parbox{6cm}{Some ad blockers like uBlock, Disconnect, Ghostery, and PrivacyBadger have good anti-tracking effectiveness against many commonly used trackers. However, they often struggle to block fingerprinting scripts and less commonly used trackers.} \\
    \midrule
    \parbox{2cm}{\centering Borgolte and \\ Feamster (2020)} & \xmark & \parbox{2.5cm}{Number of blocked
      (HTTP) resources and cookies} & $\checkmark$ & \xmark & \parbox{6cm}{Ad blockers can improve user privacy and browsing experience by reducing website load time, data transfers, and processor usage, providing additional benefits from anti-tracking effectiveness.} \\
    \midrule
    \parbox{2cm}{\centering This study} & \parbox{2.5cm}{Number of blocked ads} & \parbox{2.5cm}{Number of blocked trackers} & $\checkmark$ & $\checkmark$ & \parbox{6cm}{All four ad blockers (Adblock Plus, AdGuard, Stands, uBlock) have equal ad-blocking effectiveness in their MV3 instances compared to their MV2 counterparts. Adblock Plus and Stands block more trackers under MV3 than under MV2, while AdGuard and uBlock do not; using multiple MV3 ad blockers improves anti‑tracking relative to several single MV3 blockers.} \\
    \bottomrule
  \end{tabular}
  \label{tab:lit-review-table}
\end{table*}

\subsection{Variance and Reliability Analysis for Browser-Based Experiment}\label{appendix:variance-analysis}

In this section, we evaluate the consistency of our results across multiple runs of the browser-based experiment. Using variance analysis, we examine how much the results, specifically, the number of blocked ads and trackers, change when we repeat the runs of the browser-based experiment several times.

We use statistical measures of standard deviation (SD) and coefficient of variation (CV) to assess the consistency of the results. A low CV indicates that the results are consistent and do not vary much between runs. In addition, we compute the intraclass correlation coefficient (ICC), which measures how consistent the results are across runs by comparing the variation between different websites and ad blockers to the variation caused by repeating the browser-based experiment. Specifically, we report ICC(C,1)—a two-way, consistency, single-measure form of the intraclass correlation—together with 95\% confidence intervals. A high ICC (>0.75) indicates high consistency across runs. 

Table~\ref{tab:variance_analysis} presents the results of a variance and reliability analysis for five separate measurement runs of our browser-based experiment involving a main sample of 924 websites.

\begin{table}[H]
\caption{Variance and reliability analysis for the number of blocked ads and trackers across five runs of browser-based experiment. CV means coefficient of variation. ICC means intraclass correlation coefficient and we report its values with 95\% confidence intervals. Multiplying the 9 browser instances (Adblock Plus MV2, Adblock Plus MV3, AdGuard MV2, AdGuard MV3, Stands MV2, Stands MV3, uBlock MV2, uBlock MV3, and MV3+) with the number of websites (924) yields the number of observations (N = 8,316).}
\centering
\small
\begin{tabular}{lcc}
\toprule
\textbf{Metric} & \textbf{Blocked Ads} & \textbf{Blocked Trackers} \\
\midrule
Average SD & 0.59  & 2.70   \\
Average CV                         & 0.67 & 0.20  \\
No deviation & 4,443 (53.43\%)      & 1,581 (19.01\%)                            \\
Max-min difference $\leq$ 1       & 5,612 (67.48\%)       & 2,609 (31.37\%)       \\
\midrule
ICC [95\% CI] & 0.77 [0.77, 0.78] & 0.89 [0.88, 0.89]   \\
\midrule
N                                 & 8,316              & 8,316              \\
\bottomrule
\end{tabular}
\normalsize
\label{tab:variance_analysis}
\end{table}

Across these five runs, the results are consistent, with coefficients of variation of 0.67 and 0.20 for the number of blocked ads and trackers, respectively. For blocked ads, in over half of the cases (53.43\%), the number did not change between runs. In 67.48\% of cases, the difference between the highest and lowest number of blocked ads was only one ad. The results for blocked trackers are similarly consistent, though absolute measures are higher due to a greater mean occurrence of trackers. The ICC values confirm this impression: 0.77 for blocked ads and 0.89 for blocked trackers, indicating good and excellent consistency, respectively.

These results indicate that the ad blockers performed similarly across different runs of the browser-based experiment, yielding consistent results for our two metrics.


\section{Robustness Tests}\label{appendix:robustness}

\subsection{Effectiveness of MV3 Ad Blockers Across Different Samples}\label{appendix:stratified-samples}

To validate our individual ad blocker comparisons, we repeated our independent-sample t-test analyses on two separate stratified samples. First, we stratified by website employee count. We excluded missing values and treated each observed employee count (0, 1, 10, 100, 1,000, 100,00) as a stratum. Within each such stratum, we sampled without replacement up to a fixed number of websites. Second, we stratified by Tranco popularity rank. We recoded the "Outside Top 1m" rank to 1,000,000, removed missing values, and partitioned the resulting distribution into ten equal-frequency bins (deciles). We then sampled a fixed number of websites per decile without replacement.

In our sample stratified by website employee count, we collected data from 191 websites and 9 browser instances on October 29, 2024, yielding 1,719 observations. In our sample stratified by Tranco popularity rank, we collected data from 185 websites and 9 browser instances on October 29, 2024, yielding 1,665 observations. For both stratified samples, we performed independent-sample t-tests comparing the MV3 to MV2 ad blockers (excluding the MV3+ ad blocker group) and then compared the MV3+ ad blocker group to the MV3 ad blockers.

\begin{figure*}[h!]
    \centering
    \includegraphics[width=\linewidth]{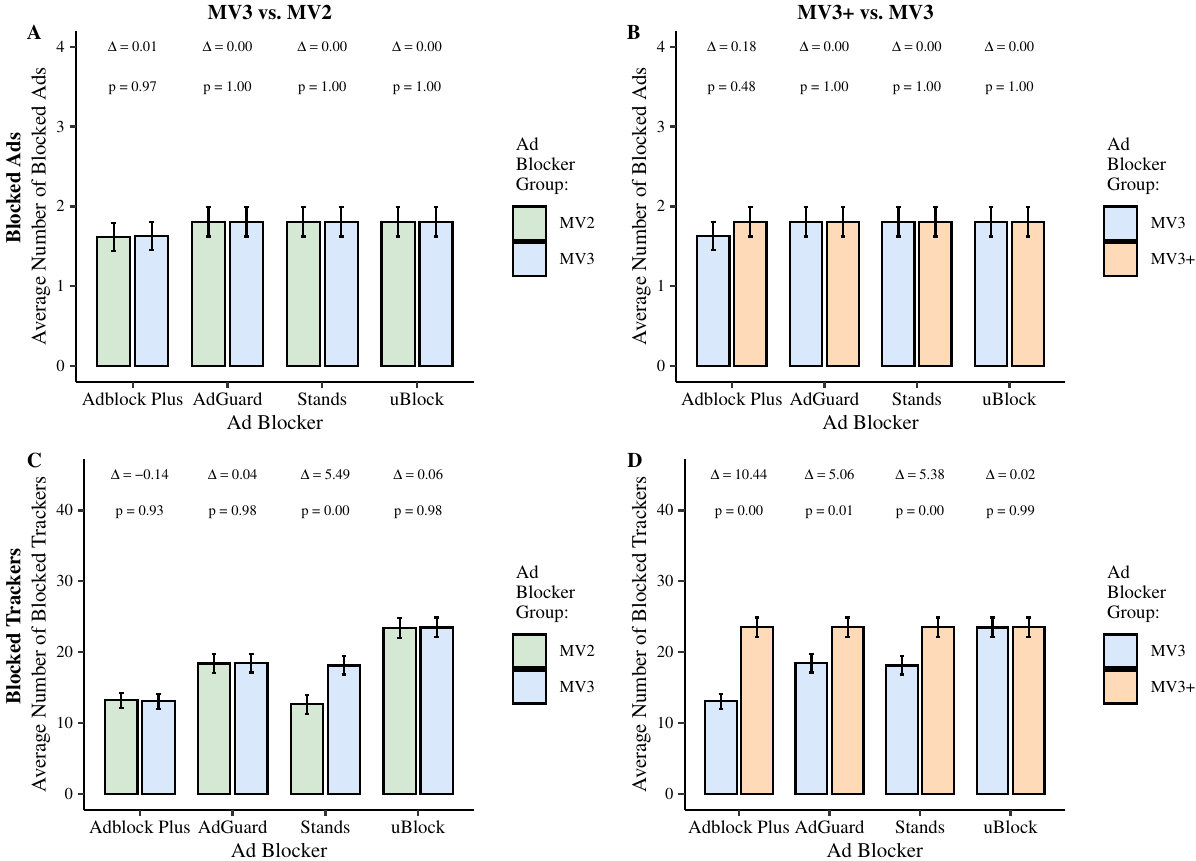}
    \caption{
          Number of blocked ads and trackers between individual ad blockers and the MV3+ ad blocker group for sample stratified by website employee count. Panels A and C show independent t-test comparisons between the mean values of the MV3 and MV2 ad blockers for the number of blocked ads and trackers, respectively. We exclude the MV3+ ad blocker group from these comparisons. Multiplying the 8 browser instances (Adblock Plus MV2, Adblock Plus MV3, AdGuard MV2, AdGuard MV3, Stands MV2, Stands MV3, uBlock MV2, uBlock MV3) with the number of websites (191) yields the number of observations (N = 1,719). Panels B and D show comparisons between the MV3+ ad blocker group and the individual MV3 ad blockers for the number of blocked ads and trackers, respectively. Multiplying the 5 browser instances (Adblock Plus MV3, AdGuard MV3, Stands MV3, uBlock MV3, and MV3+ ad blocker group) with the number of websites (191) yields the number of observations (N = 955). Values shown are from a single measurement run (no averaging); error bars represent $\pm 1$ SE across websites.
    }
    \label{fig:figure-6}
\end{figure*}

\begin{figure*}[h!]
    \centering
    \includegraphics[width=\linewidth]{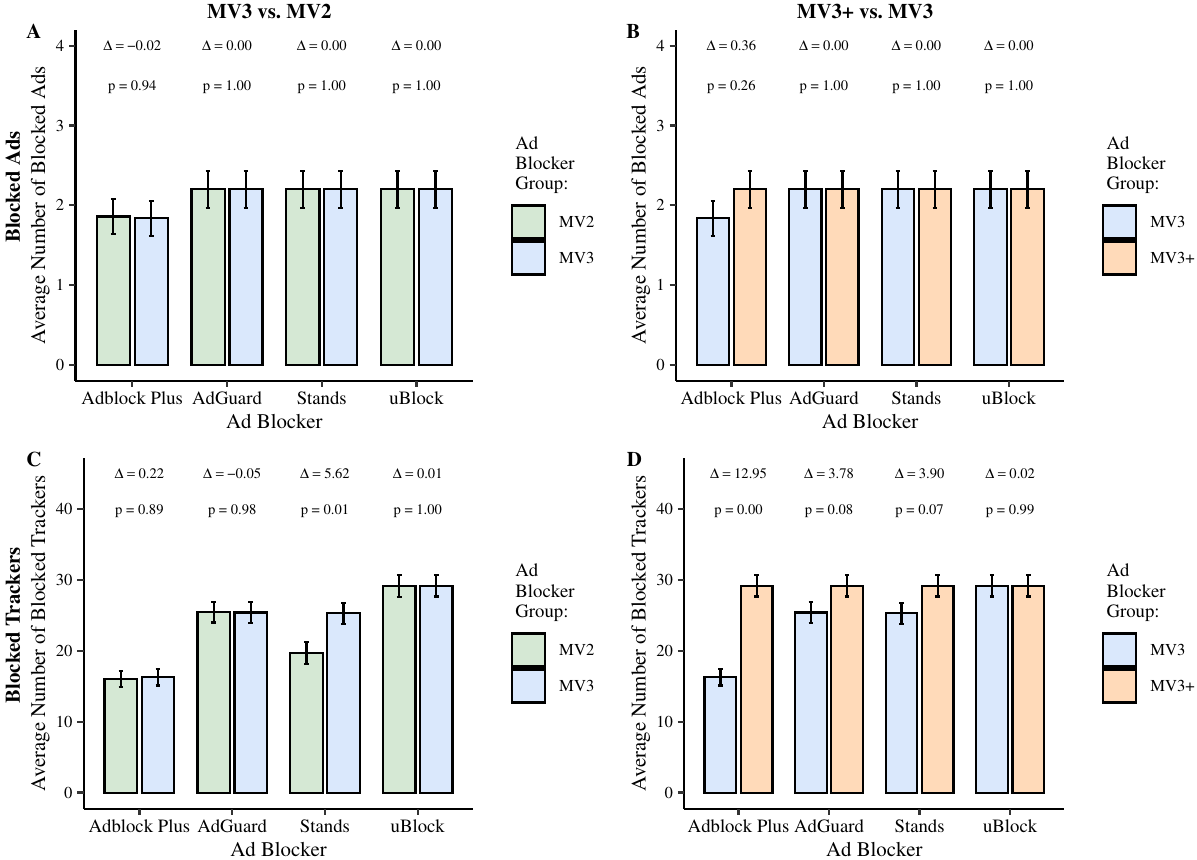}
    \caption{
    Number of blocked ads and trackers between individual ad blockers and the MV3+ ad blocker group for sample stratified by Tranco popularity rank. Panels A and C show independent t-test comparisons between the mean values of the MV3 and MV2 ad blockers for the number of blocked ads and trackers, respectively. We exclude the MV3+ ad blocker group from these comparisons. Multiplying the 8 browser instances (Adblock Plus MV2, Adblock Plus MV3, AdGuard MV2, AdGuard MV3, Stands MV2, Stands MV3, uBlock MV2, uBlock MV3) with the number of websites (185) yields the number of observations (N = 1,665). Panels B and D show comparisons between the MV3+ ad blocker group and the individual MV3 ad blockers for the number of blocked ads and trackers, respectively. Multiplying the 5 browser instances (Adblock Plus MV3, AdGuard MV3, Stands MV3, uBlock MV3, and MV3+ ad blocker group) with the number of websites (185) yields the number of observations (N = 925). Values shown are from a single measurement run (no averaging); error bars represent $\pm 1$ SE across websites.
    \label{fig:figure-7}
    }
\end{figure*}

Figures~\ref{fig:figure-6} and \ref{fig:figure-7} display the results for our samples stratified by website employee count and Tranco popularity rank, respectively. In each figure, Panels A and C show comparisons between MV3 and MV2 ad blockers for the number of blocked ads and trackers, while Panels B and D show comparisons between the MV3+ ad blocker group and the MV3 ad blockers.

For our sample stratified by website employee count, we observed no significant differences in the number of blocked ads between the MV3 and MV2 ad blockers, while for the number of blocked trackers, the Stands MV3 ad blocker outperformed its MV2 counterpart ($\Delta = 7.03$, $p < 0.001$). When comparing the MV3+ ad blocker group to the standalone MV3 ad blockers, we observed no significant differences in the number of blocked ads; however, for anti‐tracking effectiveness (i.e., the number of blocked trackers) the MV3+ ad blocker group outperformed all standalone MV3 ad blockers except uBlock MV3.

Similarly, in our sample stratified by Tranco popularity rank, the MV3 and MV2 ad blockers did not differ significantly in the number of blocked ads or trackers. Again, the MV3+ ad blocker group significantly outperformed the standalone MV3 ad blockers in blocking trackers for Adblock Plus MV3, AdGuard MV3, and Stands MV3, whereas uBlock MV3 showed no significant difference.

Together, these robustness tests across two stratified samples confirm that our findings hold despite variations in sample sizes and website selection.

\subsection{Alternative Measures for MV3 Ad Blockers Effectiveness}\label{appendix:alternative-anti-tracking}

In this robustness test, we address limitations identified in Section~\ref{sec:definition-of-two-metrics} regarding our measures for ad blocker effectiveness in our main sample of 924 websites.

We used the HTML file size (kb) of loaded websites under the usage of MV3 and MV2 ad blockers as an alternative, broader measure for ad-blocking and anti-tracking effectiveness of ad blockers. This measure follows the logic that successful blockage of unwanted ads and trackers reduces the overall size of the loaded content.

Table~\ref{tab:html-comparison} reports descriptive and t-test statistics of this robustness test. Specifically, we compared each MV3 ad blocker to its MV2 counterpart. The results of these tests increase confidence in our main findings, by showing no significant differences between most MV3 and MV2 instances of ad blockers. The only exception is the Stands ad blocker, where the MV3 instance yields a significantly lower HTML file size than the MV2 instance. These observations are consistent with our previous findings.

\begin{table}[h]
\caption{HTML file size (kb) comparison for ad‑blocking and anti‑tracking effectiveness of ad blockers. Multiplying the number of websites (924) by the two versions compared for each blocker (MV2 and MV3) yields the number of observations across both versions (1,848). Values shown are per‑website averages across five runs.}
\centering
\resizebox{0.475\textwidth}{!}{%
\begin{tabular}{lccc ccc cc}
\toprule
\textbf{Condition} &  & \multicolumn{2}{c}{\textbf{MV2}} & \multicolumn{2}{c}{\textbf{MV3}} & \multicolumn{2}{c}{\textbf{Comparison}} \\
\cmidrule(lr){3-4} \cmidrule(lr){5-6} \cmidrule(lr){7-8}
              & N (each)   & mean   & SD    & mean   & SD     & t-stat &    $p$-val \\
\midrule
Adblock Plus   & 924     & 485.95 & 501.39 & 485.23 & 510.30 & -0.03  & 0.98 \\
AdGuard        & 924     & 447.08 & 504.37 & 447.78 & 503.28 & 0.03 & 0.98 \\
Stands         & 924     & 671.67 & 506.36 & 450.28 & 505.74 & -9.40 & 0.00 \\
uBlock         & 924     & 423.12 & 470.34 & 420.65 & 466.43 & -0.11  & 0.91 \\
\bottomrule
\end{tabular}%
}
\label{tab:html-comparison}
\end{table}

To overcome shortcomings in our number of blocked trackers measure, we used the number of blocked third-party domains as an alternative, more granular measure of ad blockers’ anti-tracking effectiveness.

Unlike the number of blocked trackers, derived from third-party domains cross-referenced with the WhoTracks.me database, the number of blocked third-party domains is based directly on HTTP request data from the browser-based experiment (see Section~\ref{sec:browser-experiment}). This measure counts all distinct third-party domains blocked by each ad blocker, offering a more nuanced view of their anti-tracking effectiveness.

\begin{figure*}[h!]
    \centering
    \includegraphics[width=\linewidth]{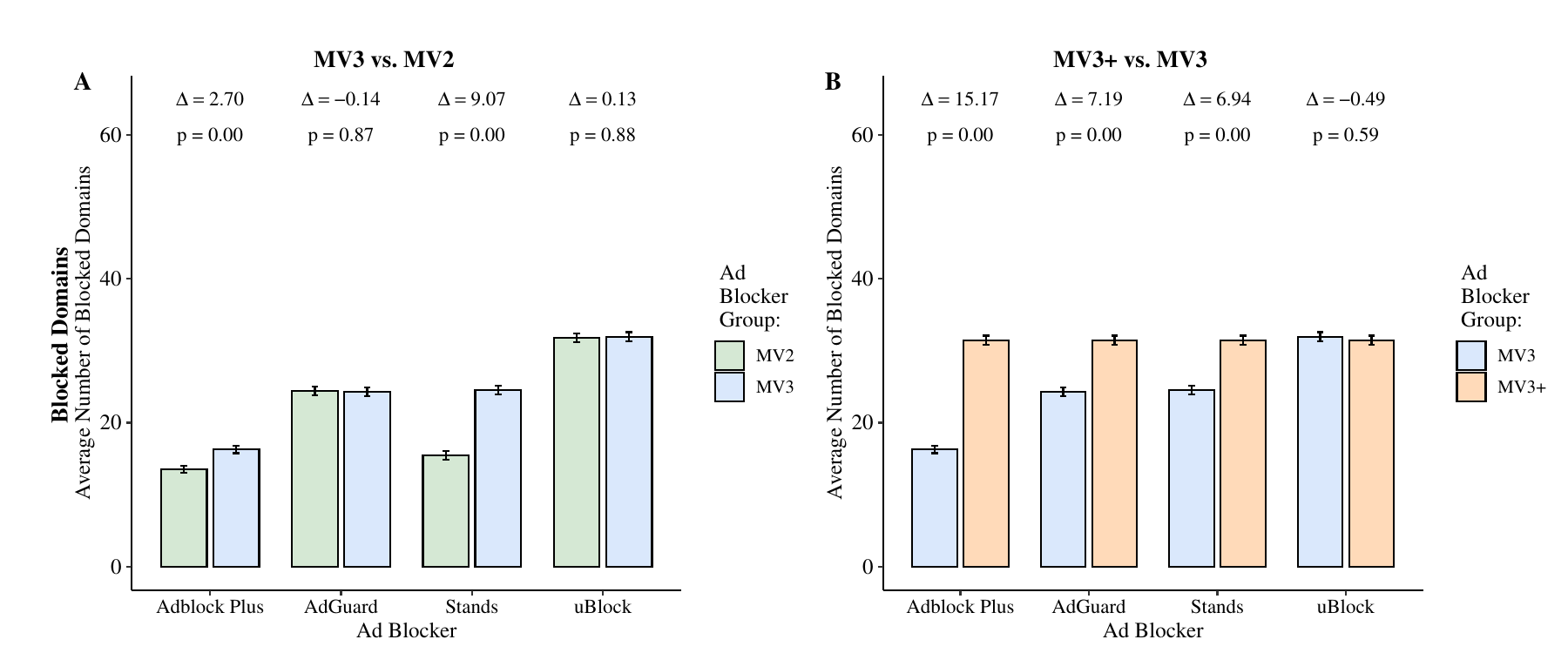}
    \caption{
        Number of blocked third-party domains between individual ad blockers and the MV3+ ad blocker group. Panel A shows independent t-test comparisons between the mean values of the MV3 and MV2 ad blockers for the number of blocked domains. We exclude the MV3+ ad blocker group from these comparisons. Multiplying the 8 browser instances (Adblock Plus MV2, Adblock Plus MV3, AdGuard MV2, AdGuard MV3, Stands MV2, Stands MV3, uBlock MV2, uBlock MV3) with the number of websites (924) yields the number of observations (N = 7,392). Panel B shows comparisons between the MV3+ ad blocker group and the individual MV3 ad blockers for the number of blocked domains. Multiplying the 5 browser instances (Adblock Plus MV3, AdGuard MV3, Stands MV3, uBlock MV3, and MV3+ ad blocker group) with the number of websites (924) yields the number of observations (N = 4,620). Values shown are per-website averages across five runs; error bars represent $\pm 1$ SE across websites.
    \label{fig:figure-8}
    }
\end{figure*}

Figure~\ref{fig:figure-8} shows independent t-test comparisons based on this alternative measure.

In Panel A, we find no significant differences between the MV3 and MV2 ad blockers in the number of blocked domains for AdGuard and uBlock; however, Adblock Plus ($\Delta = 2.70$, $p < 0.001$) and Stands ($\Delta = 9.07$, $p<0.001$) block more domains under MV3 than under MV2. In Panel B, when comparing the MV3+ ad blocker group to the standalone MV3 ad blockers, the MV3+ ad blocker group blocks more domains than Adblock Plus MV3 ($\Delta=15.17, p<0.001$), AdGuard MV3 ($\Delta=7.19$, $p<0.001$), and Stands MV3 ($\Delta=6.94$, $p<0.001$), while uBlock MV3 shows no significant difference. These results confirm a consistent pattern in anti-tracking effectiveness across both measures.

\subsection{Effectiveness of Early MV3 Ad Blockers}\label{appendix:early-mv3-ad-blockers}

This robustness test investigates whether the early versions of MV3 ad blockers were less effective immediately following their release than their later versions. It addresses concerns about potential adaptations by ad blocker providers regarding MV3’s restrictions over time.

We replicated the browser-based experiment using our main sample of 1,000 websites on July 20, 2025 using the first available versions of AdGuard MV3 (version \texttt{0.3.9}) and uBlock MV3 (version \texttt{0.1.22.806}) ad blockers. We excluded Adblock Plus and Stands ad blockers due to the availability of only a single version (Adblock Plus \texttt{4.5.1}; Stands \texttt{2.1.10}) in MV3, which we had already included in Table~\ref{tab:table-2}.

\begin{figure*}[h!]
    \centering
    \includegraphics[width=\linewidth]{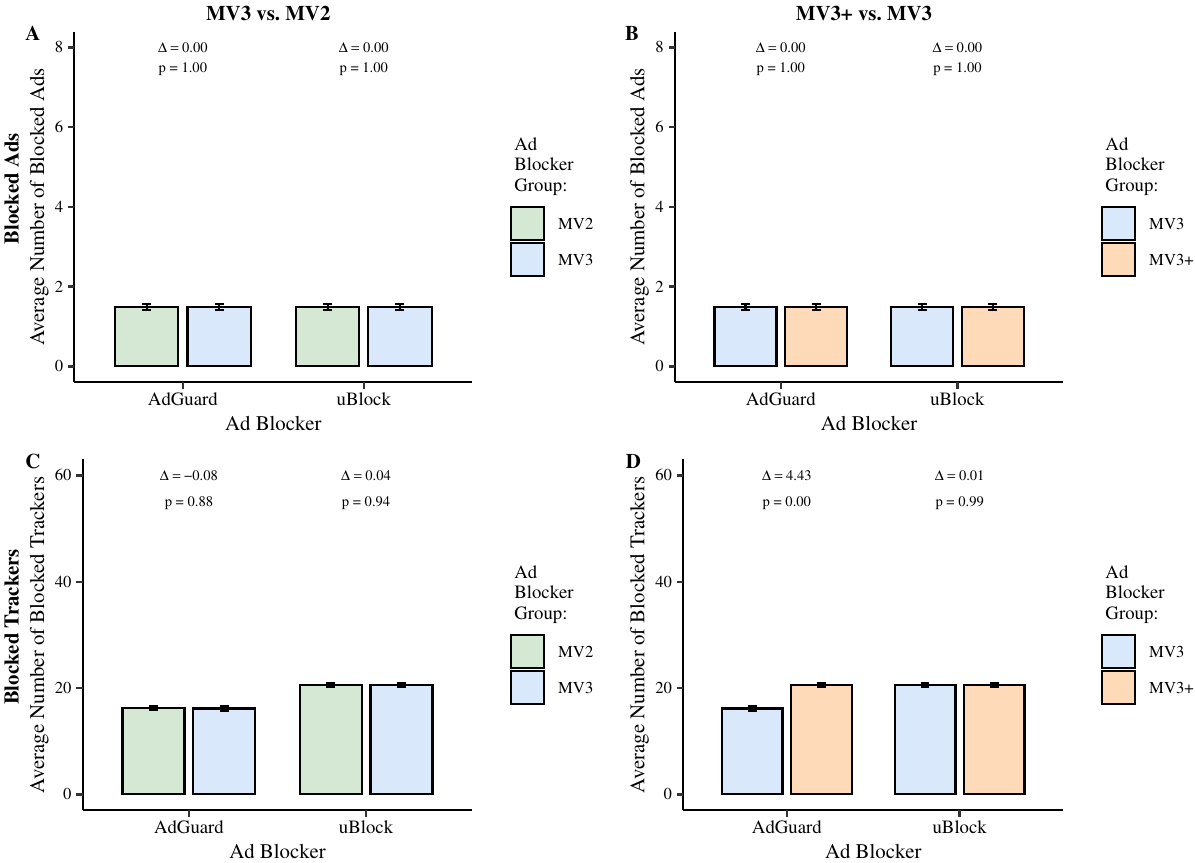}
    \caption{
    Number of blocked ads and trackers between individual ad blockers and the MV3+ ad blocker group for early MV3 ad blocker versions compared to their MV2 counterparts. Panels A and C show independent t-test comparisons between the mean values of the early MV3 ad blockers (AdGuard MV3 version \texttt{0.3.9}; uBlock MV3 version \texttt{0.1.22.806}) and their MV2 instances (AdGuard MV2 version \texttt{4.2.168}; uBlock MV2 version \texttt{1.52.0}) for the number of blocked ads and trackers, respectively. We exclude the MV3+ ad blocker group from these comparisons. Multiplying the 4 browser instances (AdGuard MV3 version \texttt{0.3.9}, uBlock MV3 version \texttt{0.1.22.806}, AdGuard MV2 version \texttt{4.2.168}; uBlock MV2 version \texttt{1.52.0}) with the number of websites (939) yields the number of observations (N = 3,756). Panels B and D show comparisons between the MV3+ ad blocker group and the individual early MV3 ad blockers for the number of blocked ads and trackers, respectively. Multiplying the 3 browser instances (AdGuard MV3 version \texttt{0.3.9}, uBlock MV3 version \texttt{0.1.22.806}, and MV3+ ad blocker group) with the number of websites (939) yields the number of observations (N = 2,817). Values shown are from a single measurement run (no averaging); error bars represent $\pm 1$ SE across websites.
    \label{fig:figure-9}
    }
\end{figure*}

This robustness test thus compares the effectiveness of these early versions against their MV2 counterparts. Figure~\ref{fig:figure-9} presents independent t-test comparisons for the number of blocked ads (Panels A and B) and blocked trackers (Panels C and D).

For the number of blocked ads, we found no significant differences between the early MV3 and MV2 versions for both AdGuard and uBlock. Similarly, for the number of blocked trackers, early MV3 ad blockers did not differ significantly from their MV2 counterparts. Moreover, in comparisons between the MV3+ ad blocker group and the early MV3 ad blockers, we observe no differences for blocked ads for both AdGuard and uBlock. For blocked trackers, the MV3+ ad blocker group blocks more than AdGuard MV3 ($\Delta = 4.43, p<0.001$), while uBlock shows no difference.

These results are consistent with those in Figure \ref{fig:figure-5}, showing that the initial MV3 ad blockers were as effective as their later versions. This consistency across different versions of MV3 ad blockers as compared to their MV2 counterparts suggests that the effectiveness of MV3 ad blockers has remained consistent over time. This outcome suggests that ad blocker providers debuted their MV3 instances of ad blockers with confidence in their effectiveness, as their core functionality did not fluctuate significantly across subsequent releases.

\subsection{Effectiveness of MV3 Ad Blockers on Firefox}\label{appendix:firefox}

A potential concern is that our Chrome-based results might be specific to that browser. To address this concern, we evaluated ad blocker effectiveness on Firefox browser to ensure cross-browser applicability of our findings.

On Firefox, we found only uBlock available in both MV3 and MV2 instances. Because Firefox does not support the Azerion Ad Expert extension, we could not count the number of blocked ads according to our methodology for Chrome (see Section~\ref{sec:browser-experiment}). For the number of blocked trackers, however, we used a custom extension to record the tracker data, mirroring our methodology for Chrome.

We initiated the Firefox browser-based experiment on a main sample of 1,000 websites across five separate measurement runs on July 12, 2025, out of which we consistently obtained results for 824 websites. We then averaged those results across five runs for our measures of interest: the HTML file size (kb) of loaded websites and the number of blocked trackers.

Table~\ref{tab:html-comparison-firefox} reports descriptive and t-test statistics for the HTML file size (kb) of loaded websites under the usage of uBlock MV3 and MV2 on Firefox. The results of this test show no significant differences between the MV3 and MV2 instances of uBlock ad blocker in HTML file size.

\begin{table}[h]
\caption{HTML file size (kb) comparison for ad‑blocking and anti‑tracking effectiveness between uBlock MV3 and MV2 on Firefox. Multiplying the 2 browser instances (uBlock MV2, uBlock MV3) with the number of websites (824) yields the number of observations (N = 1,648). Values shown are per-website averages across five runs.}
\centering
\resizebox{0.475\textwidth}{!}{%
\begin{tabular}{lccc ccc cc}
\toprule
\textbf{Condition} &  & \multicolumn{2}{c}{\textbf{MV2}} & \multicolumn{2}{c}{\textbf{MV3}} & \multicolumn{2}{c}{\textbf{Comparison}} \\
\cmidrule(lr){3-4} \cmidrule(lr){5-6} \cmidrule(lr){7-8}
              & N (each)   & mean   & SD    & mean   & SD     & t-stat &    $p$-val \\
\midrule
uBlock   & 824      & 413.47 & 428.77 & 409.66   & 421.24 & 0.18 & 0.86   \\
\bottomrule
\end{tabular}%
}
\label{tab:html-comparison-firefox}
\end{table}

\begin{figure}[h!]
    \centering
    \includegraphics[width=\columnwidth]{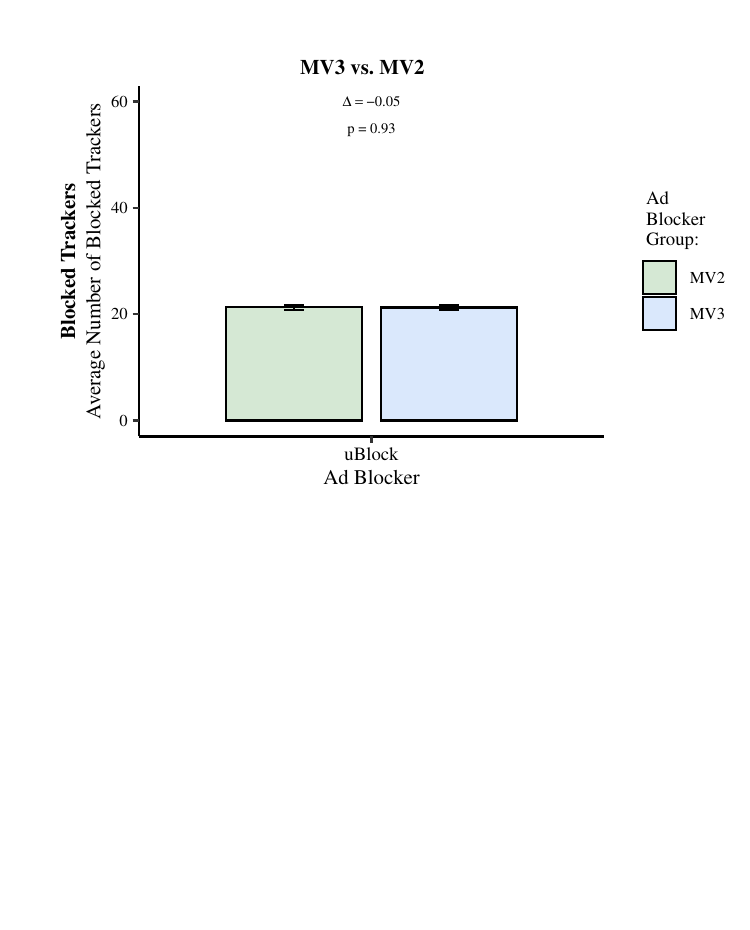}
    \caption{
    Number of trackers between uBlock MV3 and MV2 on Firefox. This figure shows independent t-test comparison between the mean values of the MV3 and MV2 instances of uBlock ad blocker for the number of blocked trackers on Firefox. We exclude the MV3+ ad blocker group from this comparison. Multiplying the 2 browser instances (uBlock MV2, uBlock MV3) with the number of websites (824) yields the number of observations (N = 1,648). Values shown are per-website averages across five runs; error bars represent $\pm 1$ SE across websites.
    }
    \label{fig:figure-10}
\end{figure}

Figure~\ref{fig:figure-10} presents independent t-test comparisons for uBlock MV3 vs. MV2 instances on Firefox. For the number of blocked trackers, we found no significant differences between the MV3 and MV2 instances ($\Delta = -0.05$, $p = 0.93$).

Overall, our Firefox results align with those on Chrome, indicating that uBlock’s effectiveness remains consistent across browsers and manifest versions.

\subsection{Visual Inspection of MV3 vs. MV2 Ad Blocker Screenshots}\label{appendix:visual_inspection}

We conducted a visual inspection to compare ad-blocking outcomes under MV3 versus MV2. From our main sample of 924 websites, we drew a random subset of 100 website visits. For each selected site, we placed the MV3 and the corresponding MV2 of the same ad blocker side by side and compared the resulting screenshots. For example, we might have randomly picked \texttt{cnn.com} and, conditional on that draw, randomly chose uBlock; we then compared the \texttt{cnn.com} screenshots produced by uBlock MV3 and uBlock MV2. The premise is that if MV3 introduces issues that are not present under MV2, systematic visual differences will emerge between the two screenshots.

As discussed in Section~\ref{sec:description-of-mv3-from-2-perspectives}, we used the following definitions in the visual assessment:

\textbf{Ad flickering}. A brief, visible appearance of ads on a page before they are hidden or removed by the ad blocker. In screenshots, ad flickering should manifest as discrepancies between visible ad counts in the images and our automated ad counts. If an ad flickers but the automation misses it, the flickering ad should sometimes still be visible in the screenshots.

\textbf{Breakage}. Unintended blocking of legitimate (non-ad) content by an ad blocker. Breakage is recorded when one manifest version blocks legitimate content that remains visible under the other manifest version.

\begin{figure}[h!]
    \centering
    \begin{subfigure}[t]{0.45\linewidth}
        \centering
        \fbox{\includegraphics[width=\linewidth]{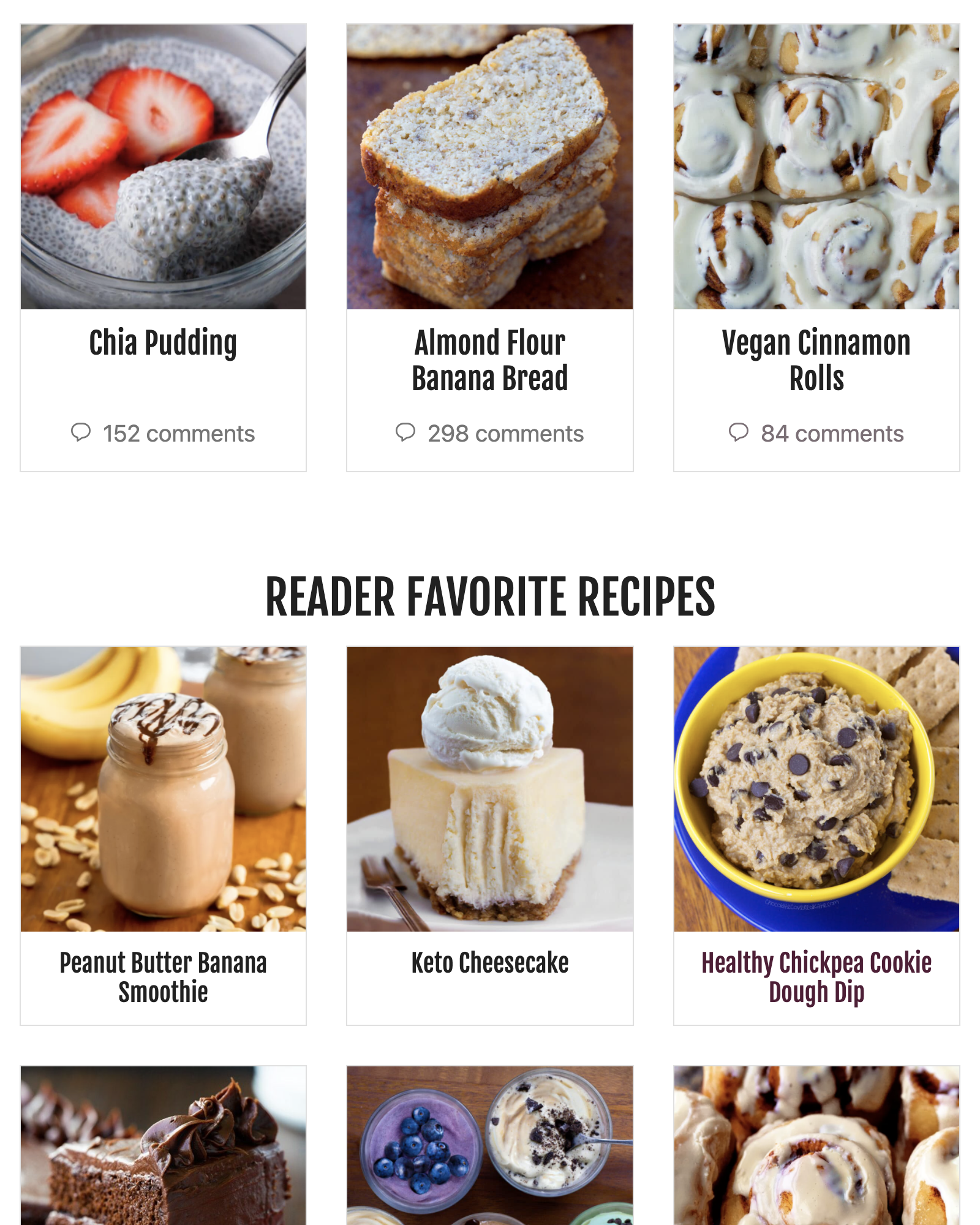}}
        \caption{Screenshot using the uBlock Origin MV2 ad blocker}
    \end{subfigure}
    \hfill
    \begin{subfigure}[t]{0.45\linewidth}
        \centering
        \fbox{\includegraphics[width=\linewidth]{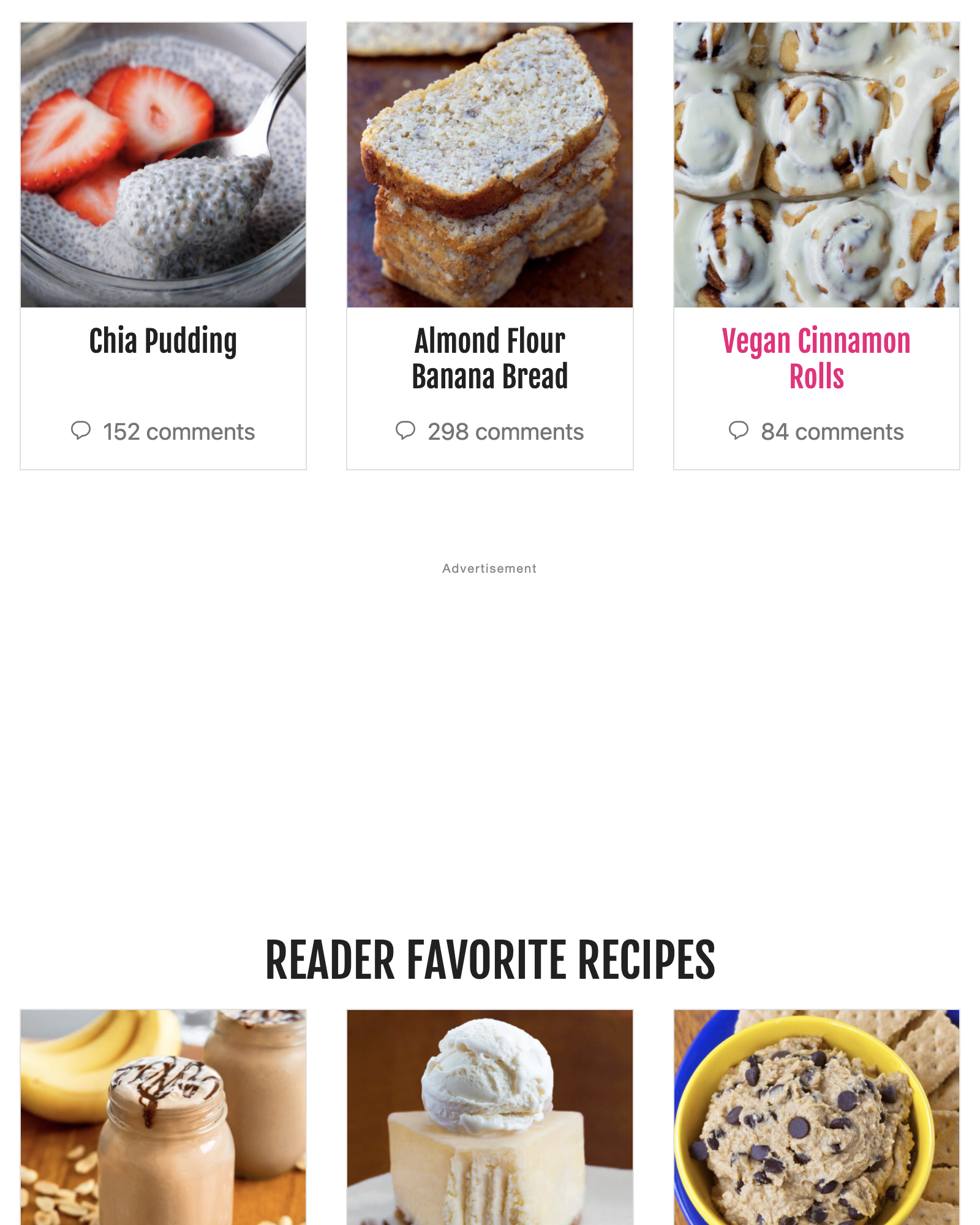}}
        \caption{Screenshot using the uBlock Origin Lite MV3 ad blocker}
    \end{subfigure}
    
    \caption{Screenshots showcasing a leftover cosmetic placeholder covering an ad under MV3, that was previously not visible under MV2, on the website \texttt{chocolatecoveredkatie.com} using uBlock Origin MV2 and uBlock Origin Lite MV3.}
    \label{fig:cosmetic-placeholder}
\end{figure}

\textbf{Cosmetic placeholders}. Residual empty spaces or visual artifacts left after ad removal (e.g., blank containers), indicating incomplete cosmetic filtering. These are recorded when such empty white spaces are visible under one manifest version but not the other. The screenshots in Figure~\ref{fig:cosmetic-placeholder} showcase one example of our comparison that reveals such a leftover cosmetic placeholder.

Two independent human evaluators assessed each outcome based on a specific question (one question per outcome), and compared the MV3 and MV2 screenshots:

\begin{itemize}
  \item ``Do you see any discrepancy in visible ads in the provided screenshots versus the provided ad counts?'' (\emph{ad flickering})
  \item ``Do you see any legitimate (non-ad) content being blocked under one manifest version that is not blocked under the other?'' (\emph{breakage})
  \item ``Do you see any empty white spaces visible under one manifest version that are not visible under the other?'' (\emph{cosmetic placeholders})
\end{itemize}

Evaluators reviewed the MV3 and MV2 screenshots side by side and answered the three questions above. We report results as Evaluator 1 (Evaluator 2). The two evaluators found no substantial ad-flickering discrepancies between MV3 and MV2: 4 (4) missed ads, split evenly across MV3 and MV2. They also observed no major breakage differences: 5 (5) MV3-only versus 5 (3) MV2-only cases. However, they found that MV3 left over cosmetic placeholders in 21\% (22\%) of cases versus 0\% (0\%) under MV2.

Within this 100-site visual audit, we did not observe meaningful MV3-specific ad flickering or systematic breakage relative to MV2. However, incomplete cosmetic filtering were notably more prevalent under MV3, with cosmetic placeholders appearing in roughly one-fifth of comparisons under MV3 and in none under MV2. Although our visual inspection is limited by the static nature of our screenshots, these findings suggest that, at present, MV3’s most visible shortcoming relative to MV2 is cosmetic rather than functional ad removal.

\end{document}